\documentclass[letterpaper,article, double-spaced]{revtex4-2}

\usepackage[dvipsnames]{xcolor}
\usepackage{graphicx}
\usepackage{ulem}
\usepackage{comment}

\newcommand{\CRA}[0]{CeRh$_{2}$As$_{2}$}

\usepackage{pdfpages}
\makeatletter
\AtBeginDocument{\let\LS@rot\@undefined}
\makeatother
\usepackage{comment}
\usepackage{amssymb}
\usepackage{amsmath}
\usepackage{graphicx}
\usepackage{tabularx}  
\usepackage{ulem}
\usepackage{float}
\usepackage{multirow}
\usepackage{threeparttable}
\newcommand\blfootnote[1]{%
	\begingroup
	\renewcommand\thefootnote{}\footnote{#1}%
	\addtocounter{footnote}{-1}%
	\endgroup
}
\usepackage[symbol]{footmisc}
\usepackage{color}
\usepackage{xcolor}

\begin{document}
	
	\title{Nodeless Superconducting State in the Presence of Zero-Field Staggered Magnetization in \CRA}

	\author{J. Juraszek}
	\affiliation{Institute of Low Temperature and Structure Research, Polish Academy of Sciences, Wrocław, 50-422, Poland}
	\author{G. Chajewski}
	\affiliation{Institute of Low Temperature and Structure Research, Polish Academy of Sciences, Wrocław, 50-422, Poland}
	\author{D. Kaczorowski}
	\affiliation{Institute of Low Temperature and Structure Research, Polish Academy of Sciences, Wrocław, 50-422, Poland}
	\author{M. Konczykowski}
	\affiliation{Laboratoire des Solides Irradi\'es, CEA/DRF/IRAMIS, \'Ecole Polytechnique, CNRS, Institut Polytechnique de Paris, Palaiseau, F-91128, France}
	\author{D.F. Agterberg}
	\affiliation{Department of Physics, University of Wisconsin–Milwaukee, Milwaukee, Wisconsin 53201, USA}
	\author{T. Cichorek}
	\affiliation{Institute of Low Temperature and Structure Research, Polish Academy of Sciences, Wrocław, 50-422, Poland}
	
	\begin{abstract}
		
		The tetragonal heavy-fermion superconductor \CRA\ with a critical temperature $T_c$\,$\approx$\,0.34\,K exhibits an intriguing magnetic field-induced transition between likely distinct superconducting states.	In zero field, an even-parity state emerges within another ordered phase of unknown origin with $T_0$\,$\approx$\,0.54\,K. Here, we investigated the spin-singlet state of \CRA\ at temperatures down to $\approx$\,0.02$T_c$ by means of local magnetization measurements performed using micro-Hall probe magnetometry. We determined the temperature dependencies of the lower critical field for both in-plane and out-of-plane field directions, and demonstrated their consistency with predominantly fully gapped superconductivity. In the magnetization measured along the $a$ axis, we found a clear increase below $T_0$, while no similar anomaly was observed along the $c$ axis. Our results place important constraints on the spin-singlet order parameter in \CRA\ and highlight an important role of static magnetic moments in the nature of $T_0$ phase.
		
	\end{abstract}

	\maketitle

	\noindent\textit{Introduction} -- Almost all known superconductors are described by a single-component wave function, where spin is the only internal degree of freedom for electrons. Since a Cooper pair is formed by pairing two spin-1/2 electrons, the spin angular momentum of a pair potential can be either spin-singlet or spin-triplet, but unequivocal experimental evidence for the latter superconductor remains elusive. However, spin-singlet and spin-triplet pair potentials can coexist in multicomponent superconductors, for which the Cooper wave functions are described by multiple complex degrees of freedom due to the presence of substantial spin-orbit coupling and crystal electric field (CEF) interactions. Candidates for multicomponent superconductors are rare, but highly sought after because one expect from them new physics such as novel topology, collective modes, and unusual response to a magnetic field \cite{Sato,Sigrist}.
	
	\vspace{1mm}
	Field-induced multiphase superconductivity appears to be realized in the recently discovered heavy-fermion superconductor \CRA\ (a critical temperature $T_c$ of approximately 0.34\,K), offering novel insights into multicomponent superconductivity \cite{Khim2021}. The material crystallizes with the tetragonal CeBe$_2$Ge$_2$-type unit cell that hosts two Ce atoms located at non-centrosymmetric sites with opposite polarity. This structural feature enables a Rashba spin-orbit coupling with alternating sign on neighboring Ce atoms layers and is believed to be crucial for highly anisotropic temperature dependencies of the upper critical field $H_{c2}(T)$ and a pronounced kink in $H_{c2}(T)$ observed in magnetic field of about 4\,T applied along the $c$ axis  \cite{Khim2021,Landaeta2022}. The latter anomaly suggests a change in the order-parameter symmetry and has been interpreted as a transition between low-field even-parity spin-singlet (SC1) and high-field odd-parity spin-triplet superconducting states. Remarkably, in fields applied with the basal plane, only the SC1 state occurs. The parity switching in \CRA\ was supported by several theoretical models, which predict different scenarios for the underlying order parameters \cite{Schertenleib2021,Nogaki2021,Mockli2021,Nogaki2022,Cavanagh2022,Suh2023,Szabo2023,Lee2024,Ishizuka2024}. 
	On the other hand, some other theoretical interpretations \cite{Machida2022,Hazra2023} as well as the results of nuclear magnetic resonance (NMR) study have challenged the spin-triplet pairing \cite{Ogata2023}.  The NMR experiments not only revealed that the spin susceptibility decreases in both superconducting states but also found in the SC1 state a clear indication of antiferromagnetism \cite{Kitagawa2022,Ogata2024}. The relevance of dipolar degrees of freedom in \CRA\ supported by zero-field nuclear quadrupole resonance (NQR) measurements \cite{Kibune2022} and the specific heat data collected in finite magnetic fields \cite{Chajewski2}. However, details of the properties of the SC1 state are still unclear. In particular, little is known about the superconducting gap structure and the possible influence of multiband effects. Certainly, the low critical temperature hinders the experimental progress in revealing the presence or absence of nodes \cite{Onishi2022,Siddiquee2023}. The difficulties are compounded by the strong sensitivity of $T_c$ to sample quality.

	In addition to its unique superconducting properties, the heavy-fermion compound \CRA\ hosts another ordered state below the temperature $T_0$\,$\approx$\,0.54\,K \cite{Khim2021}. The microscopic origin of the $T_0$ phase also remains undetermined, but likely involves higher-order multipoles of 4$f$ electrons, supported by the Ce$^{3+}$ ($J$\,=\,5/2) CEF level scheme, with two lower Kramers doublets separated by an energy of approximately 30\,K \cite{Hafner2022,Christovam2024}.  
	So far, the second order transition at $T_0$ remained invisible for magnetic measurements. Therefore, a unique case of hybridization-driven quadrupole-density wave order has been proposed to explain the nature of the $T_0$ phase, based on the fact that the estimated Kondo temperature is very similar to the CEF splitting of the quasi-quartet \cite{Hafner2022}. However, the muon-spin-relaxation ($\mu$SR) studies claimed the observation of a coherent internal field given by quasi-static magnetic moments observed below $T_0$ \cite{Khim2024}. In addition, the inelastic neutron scattering (INS) data  revealed dynamic AFM spin correlations with an anisotropic quasi-two-dimensional (2D) character, which show no discernible difference between 0.08\,K and 0.8\,K \cite{Chen2024arxiv}. Furthermore, it has recently been pointed out that the zero field $T_0$ feature of \CRA\ can be explained within the local 4$f$ approach \cite{Schmidt2024,Thalmeier2024}.

	In this Letter, we focus on the low-field magnetization of \CRA\ measured at low temperatures using Hall sensors. 
	For both in-plane and out-of-plane magnetic field $H$ directions, we found that the lower critical field $H_{c1}$ saturates below about 0.1$T_c$, indicating fully gapped superconductivity in the SC1 state.  
	Beyond that, in the temperature variation of the zero-field magnetization measured along the $a$ axis, we detected an increase that begins at $T_0$ and continues down to $T$\,$\ll$\,$T_c$.  
	Within an accuracy of our micro-Hall probe, an internal magnetic field is absent along the $c$ axis. This evidences that the low-field superconducting state of \CRA\ is created in the presence of static small-moment magnetism.

	\vspace{1.0mm}
	\noindent\textit{Experiments} --  High-quality single crystals of \CRA\ showing sharp signatures in the temperature dependence of the specific heat $C(T)$ at $T_c$\,$\simeq$\,0.33\,K and $T_0$\,$\simeq$\,0.51\,K were grown using an appropriately adapted horizontal flux growth technique as described elsewhere \cite{Chajewski1}. For details about their characterization see Supplemental Material \cite{SM} ch.\,I (SM I).
	For comparison, we have also investigated samples with lower $T_c$\,$\simeq$\,0.21\,K and with the $T_0$ anomaly barely visible (see SM VII). 
	For the local magnetization study, we used micro-Hall probe magnetometry, which can measure the magnetic field strength with micrometer resolution, thus mitigating the challenges posed by non-ellipsoidal shape of the sample under investigation. In addition, high mobility and low carrier concentration of electron gas confined in the 2D structure of a Hall sensor make it a very sensitive probe to a vertical component of magnetic field near the surface of the sample (see SM II for more details on local magnetization measurements).

	\begin{figure*}
	\begin{center}
		\includegraphics[width=\textwidth]{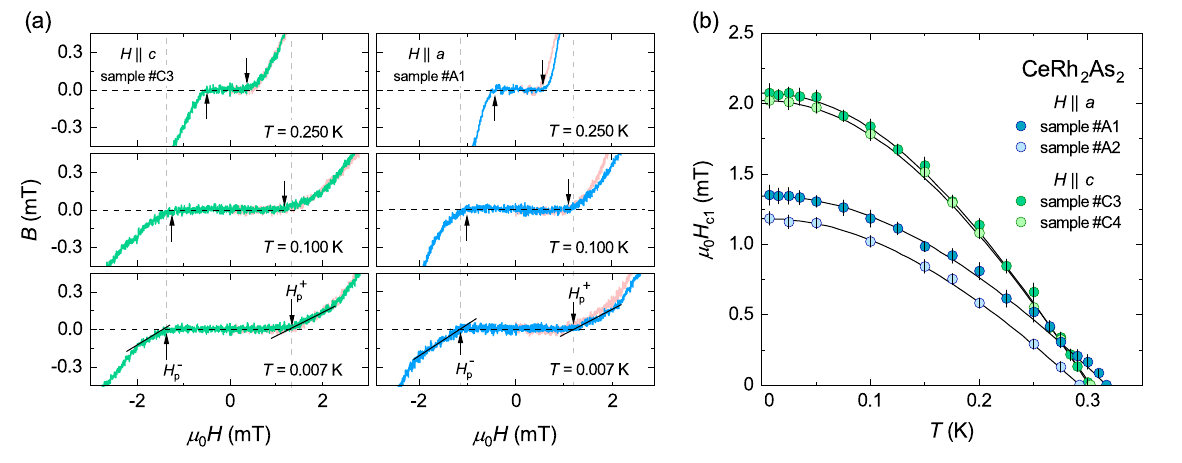}
		\caption{(a) Estimation of the field of first flux penetration $H_{\rm{p}}$ in \CRA\ at representative temperatures covering the entire superconducting state. The magnetic induction $B$ versus applied field $H$ was investigated for $H$ $\parallel$ $c$ (left panels, sample $\#$C3) and for $H$ $\parallel$ $a$ (right panels, sample $\#$A1). $B$($H$) isotherms were measured for both positive and negative magnetic field sweeps. The arrows indicate $H_{p}$'s with essentially the same value for two sweeps. The sample was zero-field cooled before each sweep. Each pink curve is the reflection antisymmetric counterpart of the curve measured for the negative sweep. (b) Temperature dependence of the lower critical field $H_{c 1}(T)$ of  \CRA\ measured for different samples both for $H$ $\parallel$ $c$ (green points) and for $H$ $\parallel$ $a$ (blue points). All sets of the $H_{c1}$($T$) data follow the conventional relation driven from the BCS theory (solid black lines).}
	\end{center}
\end{figure*}	

\vspace{1.0mm}
\noindent\textit{Results  and discussion} -- Figure 1(a) shows representative magnetic induction curves $B$($H$) of two \CRA\ samples with similar values of $T_c$,
measured for in-plane and out-of-plane field directions. The $B$($H$) isotherms at different temperatures were obtained after a shielding slope was removed from the initial part of the local magnetization (cf., SM III). When applied field reaches a critical value (indicated in Fig. 1(a) by black arrows), called the field of first flux penetration $H_{p}$, the first superconducting vortex enters the middle of the sample, and with further increase in the magnetic field strength more and more vortices penetrate the specimen. The corresponding value of $H_{c1}$ at a given temperature can be calculated from the average value of $H_{p}$\,=\,($\mid$\,$H_{p}^{+}$\,$\mid$+$\mid$\,$H_{p}^{-}$\,$\mid$)$\slash$ 2 taking into account the geometric conversion factor, where $H_{p}^{+}$ and $H_{p}^{-}$ were determined for positive and negative field sweeps, respectively (see SM III). 
In the Ginzburg-Landau theory, $H_{c1}$($T$) can be related to the normalized superfluid density as $\tilde{\rho}_{s}$($T$) = $\lambda^{2}$(0)/$\lambda^{2}$(T) $\simeq$ $H_{c1}$($T$)/$H_{c1}$(0), where $\lambda$ is the London penetration depth. In other words, the behavior of $H_{c1}$$(T)$ allows one to estimate the temperature dependence of the superfluid density, and thus the symmetry of the underlying gap function, as well as the presence of multiband effects.

Figure 1(b) displays the $H_{c1}$($T$) dependencies for in-plane and out-of-plane magnetic fields. 
In total, we measured four samples with $T_c$ varying between 0.29\,K and 0.32\,K. Special care has been taken to ensure that the Joule heating is negligible at the lowest temperatures measured (see SM IV for details). For both $H \parallel$ $a$ and $H \parallel c$, there is an apparent saturation of $H_{c1}$($T$) at $T$\,$\lesssim$\,0.1$T_c$, characteristic of fully-gaped superconductivity. Remarkably, in the entire temperature range down to about 0.007\,K, the $H_{c1}$($T$) curves can be well described by the conventional relation derived from the Bardeen-Cooper-Schrieffer (BCS) theory, suggesting that the low-field even-parity spin-singlet state in \CRA\ has an isotropic $s$-wave symmetry. 


Another important finding based on the $H_{c1}$($T$) data is a rather moderate anisotropy of the Meissner state in \CRA. We found that the mean value of $\mu_0 H_{c1}(0)$\,$\simeq$\,2.05\,mT approximated for the $c$ axis is by a factor of 1.6 larger than the corresponding value of approximately 1.25\,mT along the $a$ axis. (We note that $T_c$'s of samples $\#$A1 and $\#$A2 measured for $H\parallel a$ are somewhat different, but their averaged value of $\approx$\,0.30\,K is quite similar to $T_c$'s of samples $\#$C3 and $\#$C4 measured for $H\parallel c$.)  While the anisotropy of $H_{c1}$ is much smaller than the very large anisotropy in the upper critical field ($H_{c2}$(0)\,$\approx$\,14\,T for $H$ $\parallel$ $c$ and 1.9\,T for $H$ $\parallel$ $ab$ \cite{Khim2021}), our observation is firmly comparable with the anisotropy factor of nearly 2.5 inferred from the conventional estimation of the Pauli critical fields \cite{Khim2021,Machida2022}, which reflect the property of the SC1 state, alike the lower critical field. Further parameters characterizing the SC1 state are presented in SM V.
	
Finally, we note the lack of any clear signatures in the $H_{c1}$($T$) curves collected for \CRA\ for multigap superconductivity. It is known that $H_{c1}$($T$) is very sensitive to the presence of two (or more) distinct energy gaps, as has been shown experimentally for, e.g., the heavy fermion skutterudite PrOs$_{4}$Sb$_{2}$ \cite{Juraszek2020} and several iron-based superconductors \cite{Ren2008, Ge2012, Hafiez2013}. The absence of any observable deviation from the conventional BCS behavior supports a single-gap superconductivity in \CRA. 
	
Neverheless, assuming that the normalized superfluid density in \CRA\ can be estimated from the $H_{c1}$($T$) data, one might be inclined to describe its in-plane and out-of-plane components using a simplified two-gap $s$-wave model, and this gives reasonable agreement (see SM VIII). The fitted gap values are $\Delta_{1}$(0) = 1.9$k_{\rm{B}}T_c$ and $\Delta_{2}$(0) = 0.6$k_{\rm{B}}T_c$, with a rather minute fraction for the smaller gap of $x$\,=\,0.2\,[cf.\,Figs.\,S9(a,c)]. Interestingly, the two-gap multisymmetric spin-singlet model with $x$\,=\,0.3 for the $d$-wave gap shows reasonable agreement above 0.2$T_c$, but the agreement becomes poor at low temperatures, since for this model $\tilde{\rho}_{s}$($T$) is linear but the data are not [cf.\,Figs.\,S9(b,d)]. On the other hand, for the $s+d$ model with $x$\,$\lesssim$\,0.2, a deviation from the flat temperature dependence at low temperatures is within the resolution of our data, but such two-gap fits fail to describe the data at $T$\,$>$\,0.3$T_c$ (not shown).
	
As stated above, the most likely scenario for \CRA\ is that the in-plane and out-of-plane $H_{c1}$($T$) variations reflect an isotropic $s$-wave pairing in the SC1 state. 
Although this possibility is barely discussed in the context of \CRA\ \cite{Nogaki2021}, since it does not naturally emerge from the consideration of electronic correlations, we recall the lesson of the first heavy-fermion superconductor CeCu$_2$Si$_2$ with $T_c$\,$\approx$\,0.6\,K, for which experiments only below 0.1\,K showed fully gapped superconductivity, and thus ruled out the usual single-band $d$-wave state hypothesis \cite{Kittaka2014,Takenaka2017,Pang2017}.  Instead, a nodeless $d$-wave state has been suggested, in which the $d$-wave symmetry is attributed to a non-trivial single-electron orbital pairing \cite{Pang2017}. Such a state is possible provided that the single-particle hopping between the orbitals that pair is smaller than the superconducting gap magnitude. The origin of such a state in \CRA\ is unclear.

Another possible explanation relies on the expectation that the interband couplings can renormalize the contribution due to the $d$-wave gap to $\tilde{\rho}_{s}$($T$), and this may give a better fit to  the $s\,+\,d$ model at higher temperatures. However, it will not significantly increase the possible fraction of the $d$-wave gap, since its linear temperature dependence at low temperatures is not expected to be altered by interband coupling. Nevertheless, the $s\,+\,d$ model with a small fraction $x$\,$\lesssim$\,0.2 seems to be consistent with a $d$-wave state driven by Van-Hove hot spots \cite{Chen2024,Lee2024}. At these hot spots, the gap is not zero and it changes sign between the hot spots to give the $d$-wave symmetry. If these hot spots dominate the density of states, then any Fermi surface near the center of the Brillouin zone would have a smaller contribution to $H_{c1}$.

\begin{figure}[h]
	\begin{center}
		\includegraphics[width=80mm]{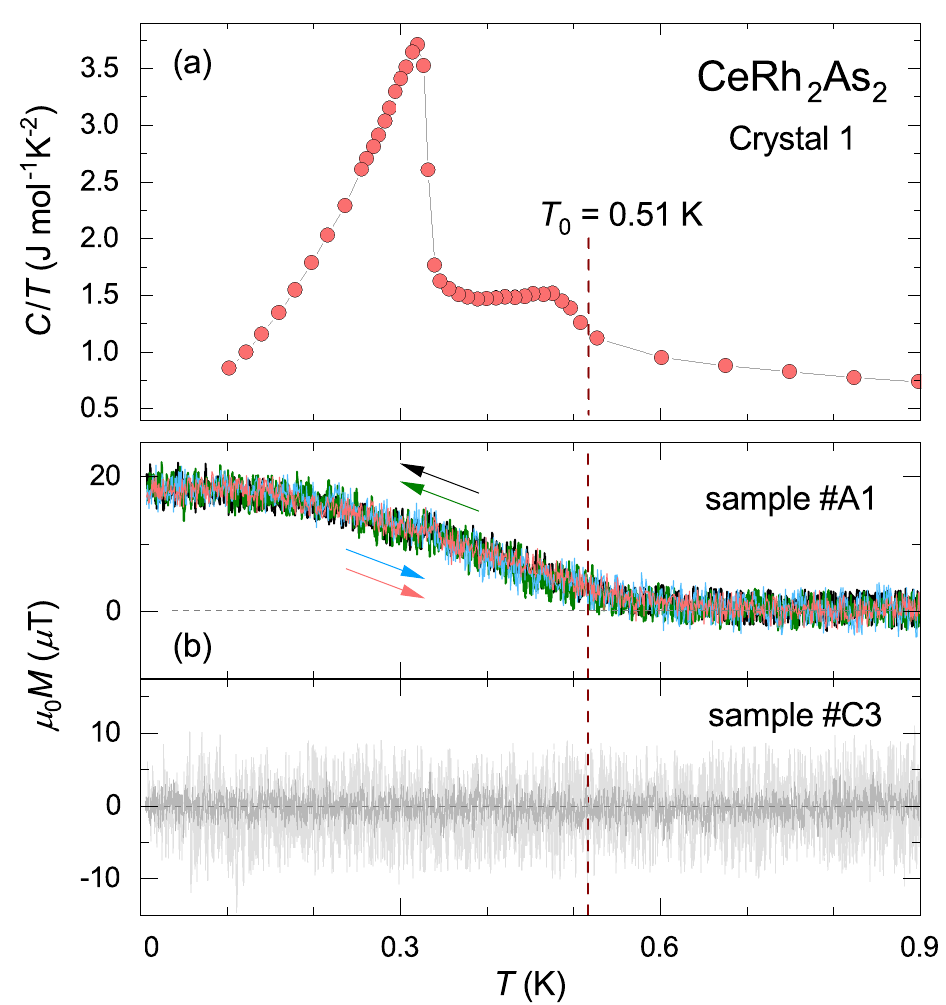}
		\caption{(a) Specific heat of \CRA, plotted as $C/T$ versus temperature, showing sharp signatures of superconductivity and undetermined phase at $T_c$ and $T_0$, respectively. From this crystal sample $\#$A1 was prepared and other single crystals, labeled 2$-$4, show similar $C/T$ data. (b) The zero-field magnetization of \CRA\ measured as a function of temperature using the Hall sensor. In these experiments, the injection current of $j$\,=\,1\,$\mu$A was used to achieve sufficient resolution by negligible heating down to 0.1\,K. Top: The $M(T)$ data taken when the $a$ axis of sample was perpendicular the active area of Hall sensor. Bottom: Analogous results for the $c$ axis. Different color intensities correspond to different injection currents (light gray $j$\,=\,1\,$\mu$A, dark gray $j$\,=\,3\,$\mu$A).}
		\label{fig:Fig1}
	\end{center}
\end{figure}

 We now turn to the question of static magnetic order below $T_0$, which is another aspect of our study of the local magnetization in \CRA.  It is worth recalling that the spectroscopic experiments, carried on large assemblies of single crystals, suggested a dynamical nature of the magnetic coupling in this compound, though their results were somewhat contradictory. Indeed, the temperature-dependent behavior (albeit measured only up to 0.6\,K) and the absolute values of the zero-field and weak transverse-field $\mu$SR time spectra are very similar, indicating comparable in-plane and out-of-plane components \cite{Khim2024}. In turn, the INS data revealed quasi-2D AFM spin fluctuations with no discernible difference between 0.08\,K and 0.8\,K \cite{Chen2024arxiv}.

Figure 2(b) shows the temperature dependence of the magnetization $M$($T$) of \CRA\ measured in zero applied field for two samples prepared from different single crystals, but with very similar $C(T)/T$ characteristics to that presented in Fig. 2(a). In the upper panel it is shown the result obtained on sample $\#$A1 that was oriented with its $a$ axis perpendicular to the active area of Hall sensor. The in-plane $M(T)$ variation shows a clear upturn that starts at about 0.55\,K from negligibly small values and continues smoothly down to about 0.15\,K, reaching an increase of 18\,$\mu$T. This behavior  was double checked on cooling and heating the specimen, and no hysteresis was found. Strikingly different result was obtained for sample $\#$C3 that was mounted on the Hall device in a manner enabling investigation of the magnetization along the $c$ axis. The out-of-plane $M(T)$ data show no distinctive features below 0.9\,K [see lower panel of Fig. 2(b)]. We emphasize that investigations of several other samples of \CRA\ confirmed the existence of a static internal field below $T_0$, although its magnitude was found to be sample dependent. It is noteworthy that the increase of in-plane $M$($T$) below $\approx$0.5\,K was also detected in a lower quality crystal with hardly discernible $T_0$ anomaly in the specific heat [see Fig. S8(a)].

\begin{figure}[h]
	\begin{center}
		\includegraphics[width=95mm]{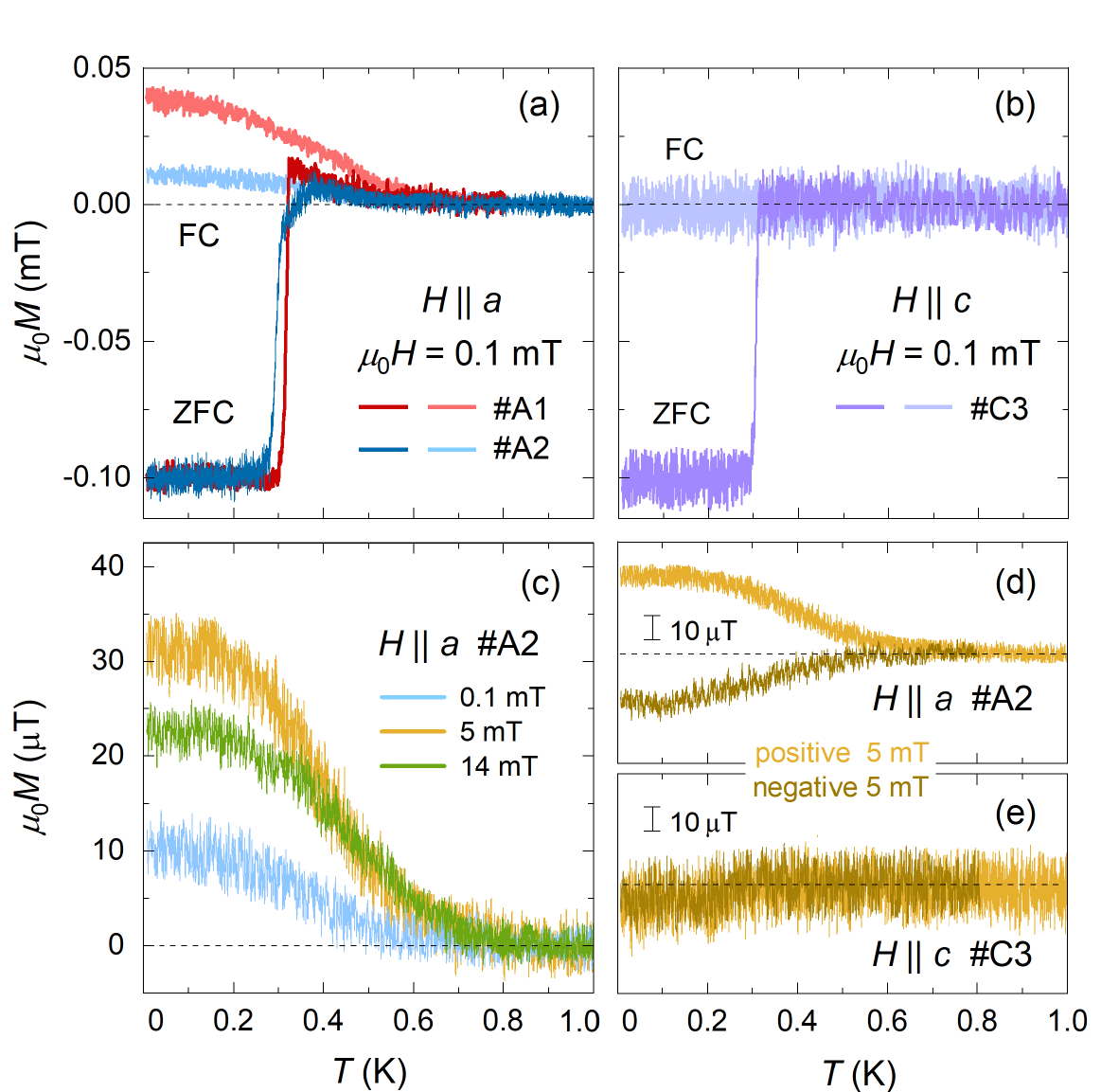}
		\caption{(a) The  in-plane magnetization under the zero-field-cooled (ZFC) and field-cooled (FC) conditions for two \CRA\ samples measured in $\mu_0H$\,=\,0.1\,mT. (b) Similar $M(T)$ results for another sample but with the $c$ axis perpendicular to the Hall sensor. (c) The in-plane magnetization data taken after cooling sample $\#$A2 in various applied magnetic fields. (d) The $M(T)$ data after cooling in a 5\,mT field with the opposite directions parallel to the $a$ axis.  (e) Corresponding $M(T)$ results for $H \parallel  c$.}
		\label{fig:Fig1}
	\end{center}
\end{figure}

Figure 3(a) compares the temperature dependencies of the zero-field cooled (ZFC) and field-cooled (FC) magnetization of samples $\#$A1 and $\#$A2 of \CRA\ measured at $T<$\,1\,K and in $\mu_0H$\,=\,0.1\,mT applied along the $a$ axis. Whereas sample $\#$A2 displays the onset of superconductivity at somewhat higher temperature of 0.38\,K, in both samples the SC1 state emerges at almost the same critical temperature, as signaled by a drop in their ZFC curves. Most importantly, for both samples, the diamagnetic signal is preceded  by an increase in $M$($T$) below $T_0$. For the FC curves, the diamagnetic signal is absent and the $M$($T$) dependencies show  the upturns similar to those shown in Fig.\,2(b). Again, it is worth noting that the FC curve taken in $H \parallel c$ was temperature invariant below 1\,K [Fig.\,3(b)]. 	

Figures 3(c,d) display the magnetic field variation of the upturn in field-cooled $M$($T$) below $T_0$, inspected for sample $\#$A2. It was found that (i) the feature becomes more pronounced in higher fields, although not monotonically, (ii) the signal reverses with the field direction, and (iii) this sign change is also observed under the ZFC condition [see Fig. S7(a)]. 
Similar findings were made also for sample $\#$A1 (not shown), but no effect of applied fields was found along the $c$ axis [Fig.\,3(e)].

Some further evidence for the presence of the in-plane magnetic signal in \CRA\ comes from the $B(H)$ isotherms shown in Fig. 1(a). Indeed, clear differences between positive and negative field sweeps were observed for $H \parallel a$ in $H$\,$>$\,$H_p$, but such differences are negligible for $H$\,$\parallel$\,$c$. In the former case, the reflection antisymmetric counterparts of the curves measured with the negative field sweeps (pink lines) distinctly differ from the curves taken with the positive field sweeps. This is in line with the $M$($T$) results depicted in Figs. 3(d,e) and Fig. S7, and emphasizes, in addition to the magnetocrystalline anisotropy, that the magnetic signal detected below $T_0$ is dependent on the magnetic field polarity.
	
The low-temperature behavior of the local magnetization measured along the $a$ axis, i.e., a clear increase in the vicinity of $T_0$, a smooth rise through $T_c$ and a flattening deep in the superconducting state, distinctly differs from both the essentially $T$-independent $\mu$SR time spectra below $T_c$ \cite{Khim2024} and the $T$-independent AFM spin fluctuations seen in the INS data below 0.8\,K \cite{Chen2024arxiv}. We note that the internal field strength measured by the Hall probe is two orders of magnitude weaker than $\approx$\,5\,mT inferred from the $\mu$SR experiment below 0.3 K. Our rough estimate of the static magnetic moment detected in the Hall magnetometry is only a few 10$^{-2}$\,$\mu_{B}$/Ce atom. Such a small value may explain the lack of clear evidence for AFM order in the neutron diffraction study that placed the upper limit of the staggered magnetization at a level of 0.3\,$\mu_B$ at temperatures down to 0.08\,K \cite{Chen2024}. Remarkably, the absence of a magnetic signal along the $c$ axis is in accord with the quasi-2D AFM spin correlations anticipated from the INS data.

Our study revealed that not only dynamic \cite{Khim2024,Chen2024arxiv} but also static magnetic properties are consistent with the zero-field magnetism that emerges in \CRA\ below $T_0$. The observed finite and strongly anisotropic magnetization (fairly sample dependent), which reverses with the direction of the applied field and occurs on the micrometer scale, is fully consistent with the scenario of imperfect cancellation in an antiferromagnet. Based on these observations, one can relate the local magnetization of the order of 10$^{-2}$\,$\mu_{B}$ to the presence in \CRA\ of AFM domains and domain walls. The region of reorientation of the AFM order parameter at the boundaries separating different domains may have very different physical properties from those of the bulk interior of the domains\,\cite{Cheong}. Therefore, we are unable to reliably infer possible magnetic structures below $T_0$. However, the observation of local in-plane magnetization appears to be in line with a recent proposal for the behavior of the $T_0$ phase under magnetic fields applied within the tetragonal plane \cite{Schmidt2024,Thalmeier2024}. The critical factor for explaining the increase of $T_0$ with the in-plane field is the field-induced coupling between the local quadrupolar order and the local magnetic moments. Symmetry constrains this local coupling to take the form $H_y m_x O_{xy}$ \cite{Schmidt2024}, where $H_y$ is the applied field, $m_x$ is the local moment, and $O_{xy}$ is the local quadrupolar moment.  If the local moment $m_x$ is the order parameter when $H_y=0$, then $T_0$ naturally increases with $H_y$ through the induced quadrupolar moment.	
	
\vspace{1.0mm}
\noindent\textit{Conclusions} -- We  brought to bear a micro-Hall probe magnetometry to test the order parameter symmetry of the low-field even-parity superconducting state in \CRA\ and to reveal a static magnetic signal emerging at $T_0$\,$>$\,$T_c$. The results of our investigation of the lower critical field indicated fully-gaped superconductivity, and are consistent with the temperature dependence derived from the BCS theory. However, within the accuracy of our measurements of $H_{c1}$($T$), we cannot exclude the possibility of two-band superconductivity with a small $d$-wave gap fraction. In the normal state below $T_0$, we found a small but clear increase in the in-plane magnetization, while no similar feature was observed in the magnetization measured along the $c$ axis. These findings imply the existence in \CRA\ of a phase with staggered magnetization, within which the even-parity superconducting state is formed.

\vspace{1.0mm}
\noindent\textit{Acknowledgments} -- We thank M. Brando for a helpful discussion. This work was supported by the Polish National Science Centre (Project OPUS-23, No. 2022/45/B/ST3/04117). The work at UWM was supported by a grant from the Simons Foundation (SFI-MPS-NFS-00006741-02, DFA).

\clearpage

\begin{titlepage}
	\begin{center}
		\vspace*{0.5cm}
		
		\Large Supplemental Material for \\
		\vspace{0.2cm}
		\LARGE	"Nodeless Superconducting State in the Presence of Zero-Field Staggered Magnetization in \CRA"
		
		\vspace{1cm}

		\Large	J. Juraszek,$^{1}$ G. Chajewski,$^{1}$  D. Kaczorowski,$^{1}$ M. Konczykowski$^{2}$, D.F.\,Agterberg,$^{3}$\,and\,T.\,Cichorek$^{1}$
		\vspace{0.5cm}
	\end{center}

	\begin{flushleft}		
		\large \textit{$^{1}$Institute of Low Temperature and Structure Research, Polish Academy of Sciences, 50-422 Wroc\l aw, Poland}
		
		\vspace{0.2cm}

		\large \textit{$^{2}$Laboratoire des Solides Irradi\'es, CEA/DRF/IRAMIS, \'Ecole Polytechnique, CNRS, Institut Polytechnique de Paris, Palaiseau, F-91128, France}
		
		\vspace{0.2cm}	
		
		\large \textit{$^{3}$Department of Physics, University of Wisconsin–Milwaukee, Milwaukee, Wisconsin 53201, USA}

		\vspace{1cm}
	\end{flushleft}

\end{titlepage}

\setcounter{figure}{0}  
\renewcommand{\thefigure}{S\arabic{figure}}
\renewcommand{\figurename}{Figure}
\renewcommand{\tablename}{Table}

\subsection{I. Single crystal growth and basic characterization }

High-quality single crystals of {CeRh$_2$As$_2$} have been grown from Bi flux using a horizontal flux growth technique. The details of the synthesis method can be found in Ref.~\cite{Chajewski1}. In addition, the lower quality sample was used for comparison purposes (see Supplemental Material ch. VII). This was obtained by a conventional flux growth method according to the recipe described by Khim {\it et al.}\,\cite{Khim2021}. The stoichiometric chemical composition of the crystals and the homogeneous distribution of elements were confirmed by electron microprobe analysis with energy dispersive X-ray spectroscopy. The crystals were oriented using a PROTO COS Laue diffraction camera and then cut and/or polished to the desired shapes and sizes.  The orientation of the samples was then checked again (see Fig. \ref{LAUE}).

All crystals used for further measurements underwent a preliminary characterization based on the specific heat measurements. The data collected on high quality crystals (labelled 1-4) showed that all samples have almost identical specific heat characteristics [as depicted in Fig.~\ref{fig1v1}(a)]. This is particularly true for the sharp signatures in the temperature dependence of the specific heat $C(T)$ at both the superconducting transition temperature $T_c$\,$\simeq$\,0.33\,K and the temperature of the second order phase transition of unknown origin $T_0$\,$\simeq$\,0.51\,K. Details on the transition temperatures and determined values of $\Delta C/T_{\rm c}$ are collected in Tab.~\ref{tab1_parameters}.

In striking contrast is the temperature dependence of the specific heat of the lower-quality crystal of {CeRh$_2$As$_2$. As shown for comparison in Fig. S2(b) (transparent blue points),  the $T_0$ anomaly is barely visible and the superconductivity sets in at much lower temperatures, accompanied by a rather broad transition width.

	\begin{figure}[h]
		\centering
		\includegraphics[width=0.65\linewidth]{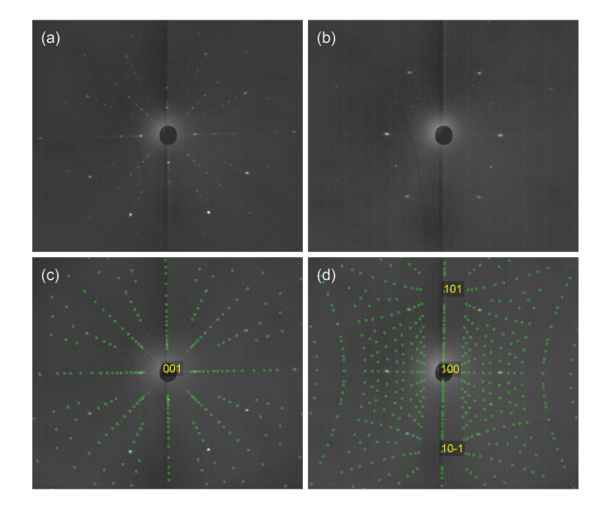}
		\label{LAUE}
		\caption{Representative Laue x-ray backscattering images collected on $\#$C3 and $\#$A1 samples of CeRh$_2$As$_2$ (raw patterns for (a) [001] and (b) [100] directions, together with corresponding overlays of theoretically generated patterns (c) and (d), respectively).} \label{LAUE}
	\end{figure}

	\begin{figure}[h]%
		\centering
		\includegraphics[width=1\linewidth]{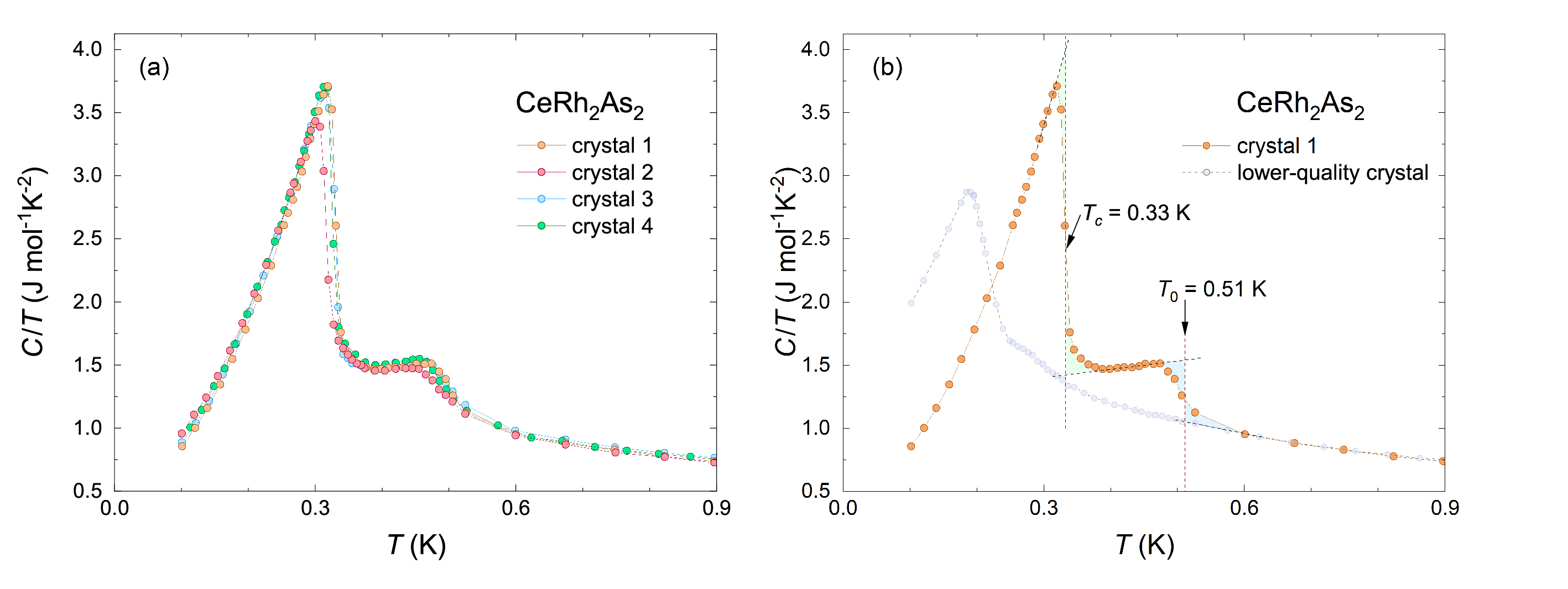}
		\caption{(a) Temperature dependence of the specific heat of CeRh$_2$As$_2$ single crystals (labelled 1-4) from which samples (labelled $\#$A1, $\#$A2, $\#$C3 and $\#$C4, respectively) were prepared for the local magnetization measurements. (b) Determination of the superconducting transition temperature $T_{c}$ and the transition temperature to the $T_{0}$ phase using the equal entropy construction, on an example of crystal 1 (orange points). The transparent blue points are the $C$/($T$) data for the lower-quality crystal.} 
		\label{fig1v1}
	\end{figure}

	\vfill
	\vfill
	\begin{table}[h]
		\caption{\label{tab1_parameters} 
			The superconducting transition temperature $T_{c}$, the superconducting transition width $\Delta T_{c}$, the transition temperature to the $T_{0}$ phase, and the size of the specific heat jump $\Delta C/T_{\rm c}$ for CeRh$_2$As$_2$ single crystals used in this study.}
		\centering
		\begin{tabularx}{80mm} {l >{\centering\arraybackslash}X >{\centering\arraybackslash}X >{\centering\arraybackslash}X >{\centering\arraybackslash}c }
			\hline \hline
			Crystal \rule{0pt}{2.6ex} & $T_{\rm c}$ & $\Delta T_{\rm c}$ & $T_{0}$ & $\Delta C/T_{\rm c}$ \\
			& [K] & [K] & [K] & [J mol$^{-1}$ K$^{-2}$]   \\
			
			
			\hline
			1  \rule{0pt}{2.6ex}  & 0.33(02) & 0.07(02) & 0.51(02) & 2.57(4) \\
			2 & 0.32(04) & 0.09(05) & 0.50(03) & 2.42(3)  \\
			3 & 0.33(02) & 0.09(04) & 0.51(02) & 2.58(4) \\
			4  & 0.33(03) & 0.08(04) & 0.51(03) & 2.57(4) \\
			\hline \hline
		\end{tabularx}
	\end{table}

	\newpage
	\clearpage

	\subsection{II. Advantages of micro-Hall-probe magnetometry}

	Micro-Hall probe magnetometry allows accurate detection of the vector magnetic field with subgauss sensitivity and micrometer spatial resolution. This technique was successfully used, e.g., for the study of atomic layers of ferromagnetic crystals \cite{Kim2019}, magnetic properties of nanoparticles \cite{Min2017}, 
	or to detect weak magnetic moments and determine their spatial distribution \cite{Li2019} [cf. Fig. S3(c)]. Another important aspect of the micro-Hall probe magnetometry is related to a study of the Meissner-Ochsenfeld effect \cite{Okazaki2010,Abdel2013,Cieplak2013,Yamashita2017,Collignon2017,Takenaka2017,Juraszek2020,Ishihara2023,Juraszek2024} because this technique minimizes disadvantages caused by 1) existence of various surface barriers that inhibit penetration of magnetic field, hence leading to overestimation of the lower critical field $H_{c1}$, and 2) distortion of magnetic field around non-ellipsoidal sample that brings about underestimation of $H_{c1}$ [cf. Fig. S3(b)].

	\vspace{-2mm}

	\begin{figure*} [h]
		\centering
		\includegraphics[width=96mm,keepaspectratio=true]{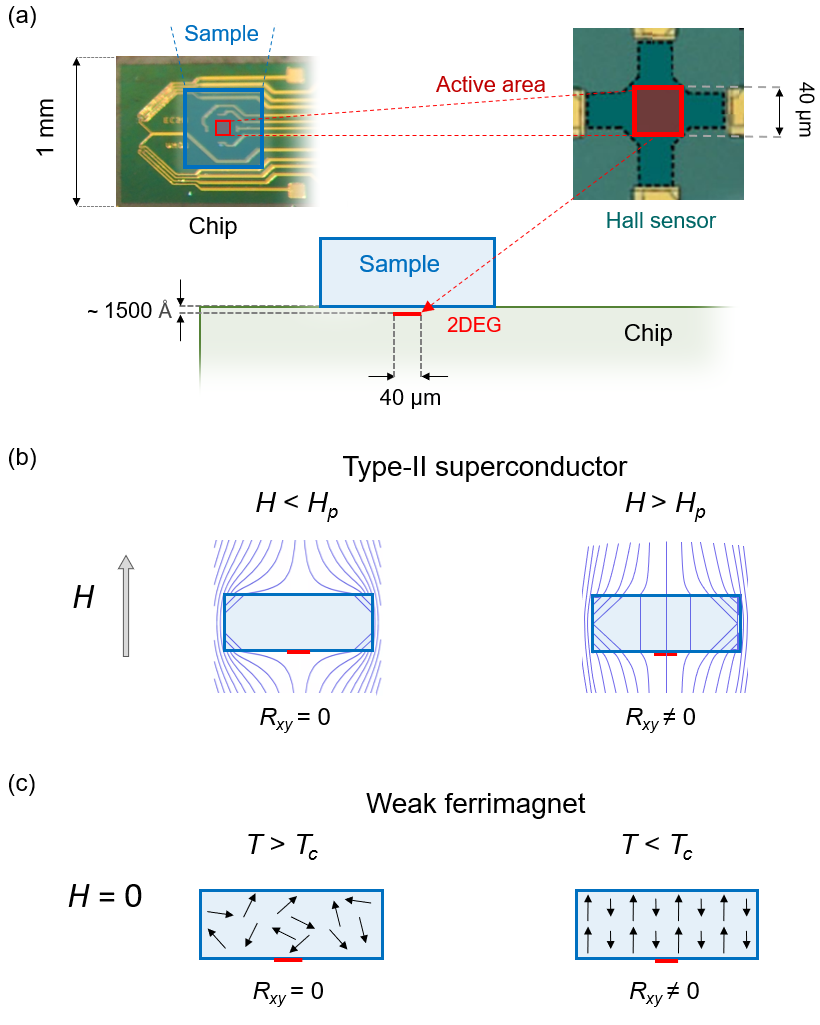}
		\vspace{-2mm}
		\caption{\textbf{Micro-Hall probe magnetometry.} (a) (top) Photograph of a chip containing the Hall sensor with an active area of 40\,$\times$\,40 $\mu$m$^{2}$ (red), which is much smaller than the cross-section of a typical sample (250\,$\times$\,250\,$\mu$m$^{2}$) (blue). (bottom) In our experiments, the centre of the sample was positioned just above the active element.  (b) Determination of the Meissner state in a type II superconductor. (c) Detection of the transition from the paramagnetic state to the magnetically ordered state with a non-zero vector magnetic field on an example of a weak ferrimagnet.}
		\label{fig:Hall0}
	\end{figure*}
	
	\vspace{-2mm}
	Hall sensors are mainly realized in a two-dimensional electron gas (2DEG) confined in GaAs/AlGaAs heterostructure. Due to a low charge-carrier concentration of about 6$\times$10$^{11}$ cm$^{-2}$ and a very high mobility of the order of 10$^{5}$ cm$^{2}$V$^{-1}$s$^{-1}$,  a 2DEG provides a large and easily measurable Hall response. Since the charge carriers move ballistically inside the 2DEG junction, the Hall resistance $R_{xy}$ is a linear function of an applied field $H$ with a sensitivity as high as 0.86 $\Omega$/mT in the wide field range up to quantization of the Hall effect. By measuring $R_{xy}$, one can accurately estimate a local magnetic induction $B_{loc}$ at the position of the active area, i.e., very close to the surface of the sample.

	The photograph in Fig. S3(a) shows a chip containing a Hall sensor that we used to locally probe the magnetization of \CRA. An active area of the Hall sensor of 40$\times$40\,$\mu$m${^2}$ is much smaller than an approximate cross-section of a typical sample of 250$\times$250 $\mu$m${^2}$. Each sample,  with a thickness of about 100 $\mu$m, was mounted on top of a chip above the active area of the Hall sensor. As a result, the active element is located about 1500\,\AA\ from the sample surface. In the experiments we carried out, the Hall sensors were placed close to the centre of the samples.

		\newpage

		\subsection{III. The field of first flux penetration and  corresponding  $H_{c1}$ value}

		Figure \ref{fig:Hp}(a) shows the initial parts of the magnetization curves $\mu_{0}M$\,=\,$B_{loc}$–$\mu_{0}H$ (where $\mu_{0}$ = 4$\pi$\,$\times$\,10$^{-7}$ H$/$m is permeability of the free space) of sample $\#$C4 measured at 0.05\,K as an example. The initial slope of the magnetization $M$/$H$ = $\alpha$ shows a nearly perfect linear dependence with $\alpha$= -0.9 slightly smaller than unity. This is due to a very small but not zero distance between the active area of the Hall sensor and the sample surface, which results in a magnetic field leakage around the sample edge. Figure \ref{fig:Hp}(b) presents a magnetic induction curve $B$($H$) that has been obtained after removing linear term (1 + $\alpha$)$\mu_{0}H$ from $B_{loc}$.} As magnetic field increases, the $B(H)$ behavior deviates from the horizontal dependence at critical value, called the field of first flux penetration $H_{p}$. A\,ﬁeld-sweep rate as small as 50 $\mu$Tmin$^{-1}$ was applied to ensure a high accuracy. We note that although the Earth's magnetic field is not completely screened at the sample location, its impact on the $B$($H$) isotherms is negligible. This is illustrated by essentially the same absolute value of $H_{p}$ for the positive and negative field field sweeps, as shown in Fig. \ref{fig:Hp}(b). The sample was zero field cooled before each sweep.

	\begin{figure*} [h]
		\centering
		\includegraphics[width=140mm,keepaspectratio=true]{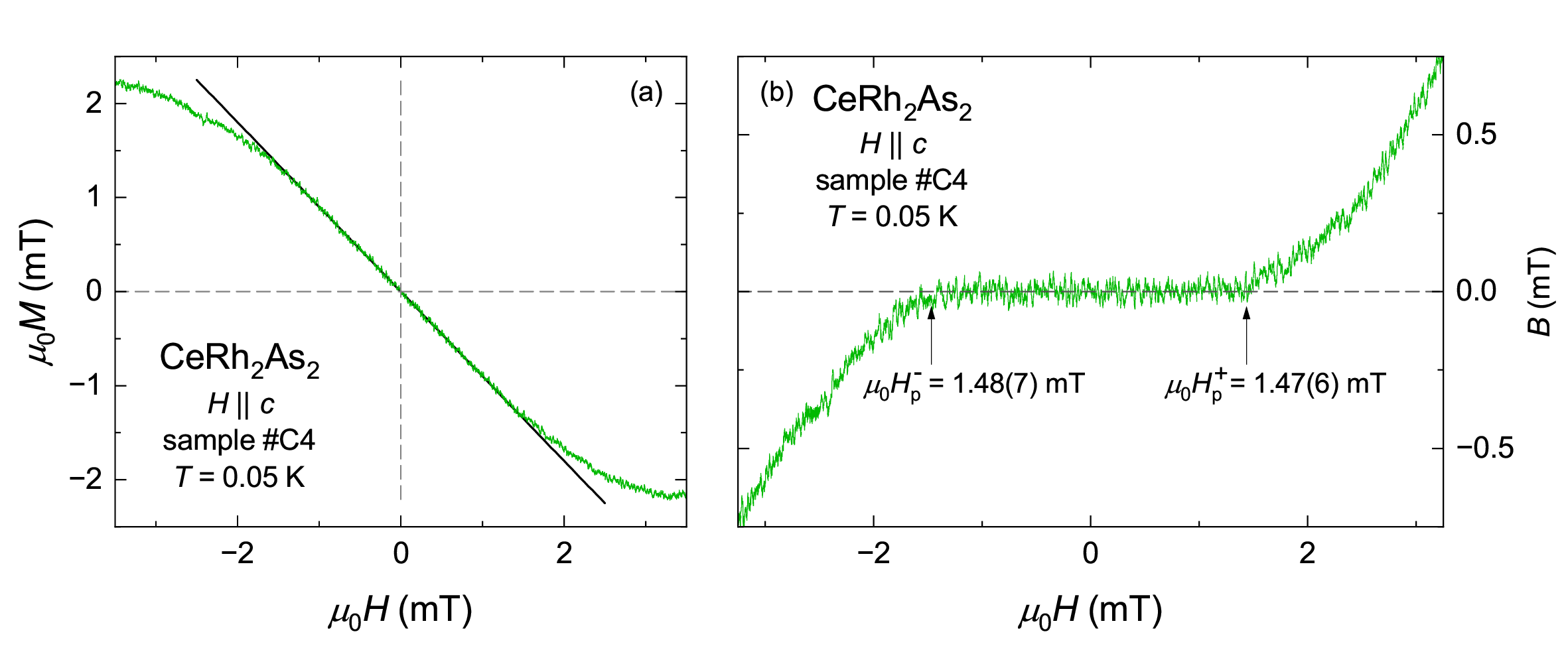}
		\caption{
			\textbf{Estimation of the field of first flux penetration.} (a) Initial part of local magnetization isotherm for sample $\#$C4 at 0.05\,K. The dashed line marks a shielding slope determined by a linear fit between -1 and 1 mT. (b) The magnetic induction versus applied field for both positive and negative magnetic field sweeps after subtraction of the Meissner slope from the $M$($H$) data shown in (a). The arrows indicate $H_p$'s.}
		\label{fig:Hp}
	\end{figure*}

	The $H_{c1}$ value at a given temperature was calculated from the average value of  $H_{p}$\,=\,($\mid$$H_{p}^{+}$$\mid$+$\mid$$H_{p}^{-}$$\mid$)$\slash$2 by taking into account the geometric conversion factor. This can be calculated for a cuboid sample with dimensions 2$m$ $\times$ 2$n$ $\times$ 2$k$ using relation \cite{Joshi2019}:
	\vspace{-2mm}
	\begin{equation}
		\label{eq:N}
		H_{c1}=\frac{H_{\rm{p}}}{1+\chi N},
	\end{equation}
	where $\chi$ is the dimensionless intrinsic magnetic susceptibility of the material in the superconducting state, which can be taken to be equal to -1, and $N$\,=[\,$ 1+\frac{3}{4}\frac{k}{m}\left ( 1+\frac{m}{n}\right )]^{-1}$ stands for the effective demagnetization factor. Because the Hall sensor probes the onset of magnetic flux penetration into the sample locally, the parameters $2m$\,=\,40\,$\mu$m and $2n$\,=\,40\,$\mu$m correspond to the width and length of its active area. The parameter $2k$\,=\,50\,$\mu$m is the dimension of the sample parallel to the magnetic field. As a result for sample $\#$C4, we obtained $N$\,=\,0.25 and $\mu_{0}H_{c1}$\,=\,2.02(7)\,mT at\,0.05\,K. 
	

	\vfill
	\clearpage
	\newpage
	\subsection{IV. Intrinsic origin of the saturation of $H_{c1}$($T$) in the $T$\,=\,0 limit.}
	
	\vspace{-1mm}
	The BCS-type characteristic of $H_{c1}$$(T)$ for \CRA\ does not allow a definitive exclusion of Joule heating below about 0.030 K, where saturation occurs. However, sample heating can be ruled out on the basis of our unpublished but previously presented data for the putative chiral superconductors PrOs$_4$Sb$_{12}$ and 4Hb-TaS$_2$\,$^{(a)}$\blfootnote{$^{(a)}$T. Cichorek, Lower critical field of the multiband superconductor PrOs$_{4}$Sb$_{12}$ with broken time-reversal symmetry. The International Conference on Solid Compounds of Transition Elements, Wrocław, Poland 2021 (oral).}$^{(b)}$\blfootnote{$^{(b)}$T. Cichorek, Vortex penetration into the putative chiral superconductor 4Hb-TaS$_{2}$ by local magnetization measurements, International Conference on Superconductivity and Magnetism, Mugla, Turkey 2024 (invited talk).}. Their $H_{c1}$$(T)$ characteristics were investigated for $j$\,=\, 0.1\,$\mu$A and 0.3\,$\mu$A, and we measured the  non-saturating dependencies down to the base temperature of 0.007\,K, demonstrating negligible heating under these conditions.
	
	\vspace{-2mm}
	\begin{figure*}[h]
		\centering
		\includegraphics[width=125mm,keepaspectratio=true]{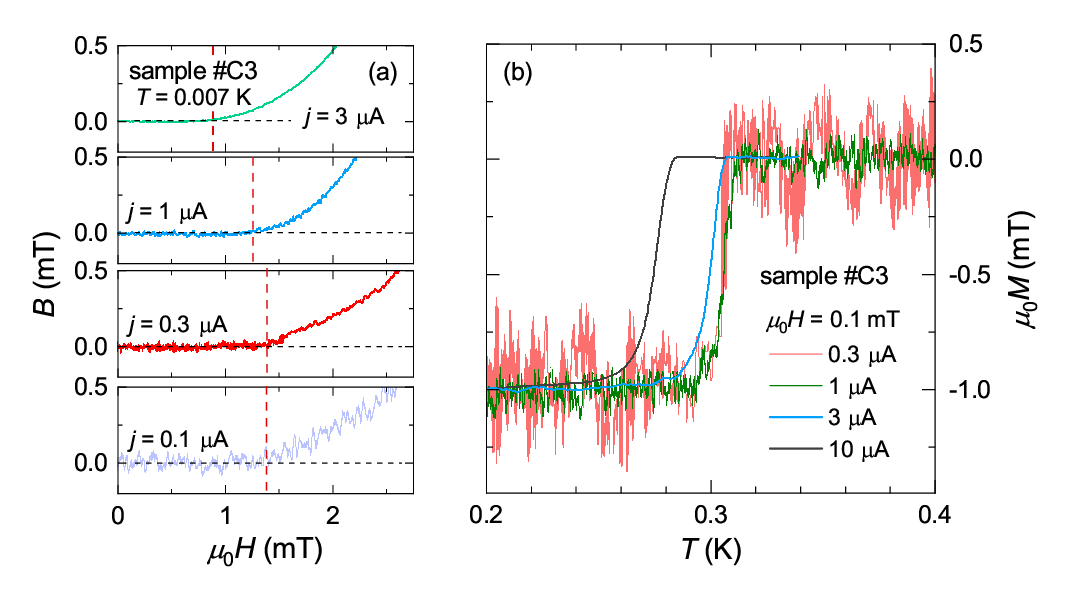}\vspace{-2mm}
		\caption{(a) Different $B$($H$) isotherms for sample $\#$C3 measured at the base temperature of 0.007 K for injection currents ranging between 0.1\,$\mu$A and 3\,$\mu$A. In each panel, the vertical dashed  red line marks the field of first flux penetration $H_p$. Note the same value of $H_p$ for $j$\,=\, 0.1\,$\mu$A and 0.3\,$\mu$A. (b) The zero-field-cooling magnetization in the vicinity of $T_c$ for injection currents ranging between 0.3\,$\mu$A and 10\,$\mu$A. Note the sharp drop in the magnetization at the same temperature for $j$\,=\, 0.3\,$\mu$A and 1\,$\mu$A.}
		\label{fig:grzanie}
	\end{figure*}

	\begin{figure*}[h]
		\centering
		\includegraphics[width=150mm,keepaspectratio=true]{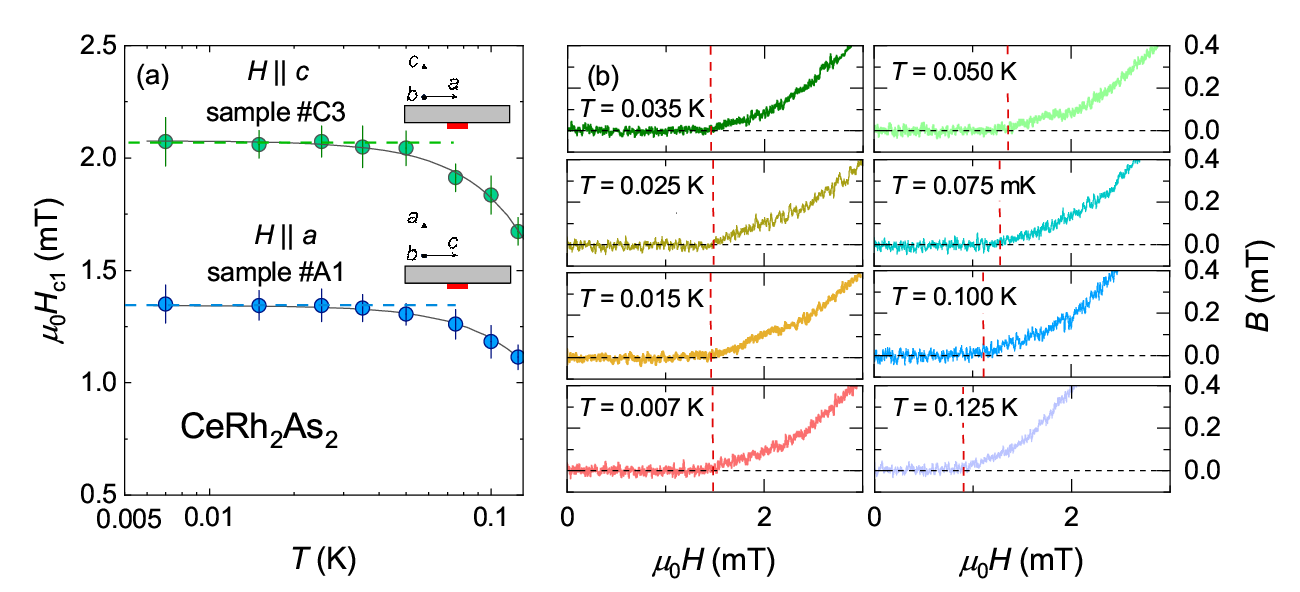}
		\caption{(a) Temperature dependence of the lower critical field for samples $\#$C3 and $\#$A1 below 0.130 K. The logarithmic temperature scale shows more detail in the temperature range where the $H_{c1}$ curves saturate (horizontal dashed lines). (b) The $B$($H$) isotherms at different temperatures from which the $H_{c1}$($T$) data for sample $\#$C3.(a) were obtained. In each panel, the vertical dashed red line marks the field of first flux penetration $H_p$.}
		\label{fig:LogT}
	\end{figure*}

	\newpage

	\subsection{V. Parameters characterizing the SC1 derived from lower critical field measurements.}
	
	In the Ginzburg-Landau theory, the lower critical field for a tetragonal superconductor is given by \cite{Tinkham}:
	
	\begin{equation}
		H_{c1}^{a}=\frac{\Phi_0}{8 \pi \mu_0 \lambda_{a} \lambda_c}\left[2 \ln \left(\sqrt\frac{\lambda_{a}\lambda_{c}}{\xi_{a}\xi_{c}}\right)+1\right],
		\label{eq:Hc1ab}
	\end{equation}
	
	\begin{equation}
		\begin{aligned}
			H_{{c1}}^{c}=\frac{\Phi_0}{8 \pi \mu_0 \lambda_{a}^2}\left[2 \ln \left(\frac{\lambda_{a}}{\xi_{a}}\right)+1\right], \\
		\end{aligned}
		\label{eq:Hc1c}
	\end{equation}
	
	where $H_{c1}^{a}$ and $H_{c1}^{c}$ are the lower critical fields for $H$\,$\parallel$\,$a$ and for $H$\,$\parallel$\,$c$, respectively. The parameters $\lambda_{a}$ and $\lambda_{c}$ are the in-plane and out-of-plane penetration depths, and  $\xi_{a}$ and  $\xi_{c}$ are the in-plane and out-of-plane coherence lengths, respectively.
	
	As shown in the main text [Fig. 1(b)], the mean value of $\mu_0 H_{c1}(0)$\,=\,1.25(7)\,mT measured for the $a$ axis is by a factor of 0.6 smaller than the corresponding value of 2.05(6)\,mT along the $c$ axis. Based on this,  we can estimate the absolute values of  $\lambda_{a}$(0)\,=\,660(12) nm and $\lambda_{c}$(0)\,=\,915(50) nm. Here, we have assumed $\xi_{a}$(0)\,=\,4.9\,nm and $\xi_{c}$(0) =\,35.4\,nm, which values were previously from the upper critical field measurements \cite{Khim2021}. 
	
	The corresponding Ginzburg-Landau parameters for the in-plane and out-of-plane directions can be estimated from:
	
	\begin{equation}
		\kappa^{a}=\sqrt\frac{\lambda_{a}\lambda_{c}}{\xi_{a}\xi_{c}},
	\end{equation}

	\begin{equation}
		\kappa^{c}=\frac{\lambda_{a}}{\xi_{a}},
	\end{equation}

	giving rise to $\kappa^{a}$\,=\,59(3) and $\kappa^{c}$\,=\,135(2) ($\kappa^{c}$=\,68(3) if the Pauli limit is considered.)  Our estimates of $\kappa^{c}$ are close to $\kappa^{c}$\,=\,100, which was roughly assessed from the thermodynamic critical field $H_{c}$ and taking into account the heavy-fermion nature of \CRA\,\cite{Szabo2023}, but clearly different from $\kappa^{c}$ = 319, which was obtained from the Knight shift decrease due to the superconducting diamagnetic shielding effect \cite{Ogata2023}.

	Finally, $H_{c1}$($T$) can be related to the temperature dependence of the normalized superfluid density as $\tilde{\rho}_{s}$($T$) = $\lambda^{2}$(0)/$\lambda^{2}$(T) $\simeq$ $H_{c1}$($T$)/$H_{c1}$(0). Assuming that the dimensionless parameter $\kappa(T)$ is effectively constant under the logarithm for the tetragonal material \CRA, we used  the experimentally determined curves of $H_{c1}^{a}$(T) and $H_{c1}^{c}$(T) to estimate the associated in-plane and out-of-plane components of the superfluid density, i.e.,  $\tilde{\rho}_{s}^{a}$($T$)\,=\, $\lambda^{2}_{a}$(0)/$\lambda^{2}_{ab}$($T$) and $\tilde{\rho}_{s}^{c}$($T$)\,=\, $\lambda^{2}_{c}$(0)/$\lambda^{2}_{c}$($T$), respectively. The results are shown in Fig. S9.

	\vspace{5mm}

	\newpage
	\subsection{VI. Sign reversal of the in-plane magnetization at low magnetic fields.}
	
	\begin{figure*}[h]
		\centering
		\includegraphics[width=160mm,keepaspectratio=true]{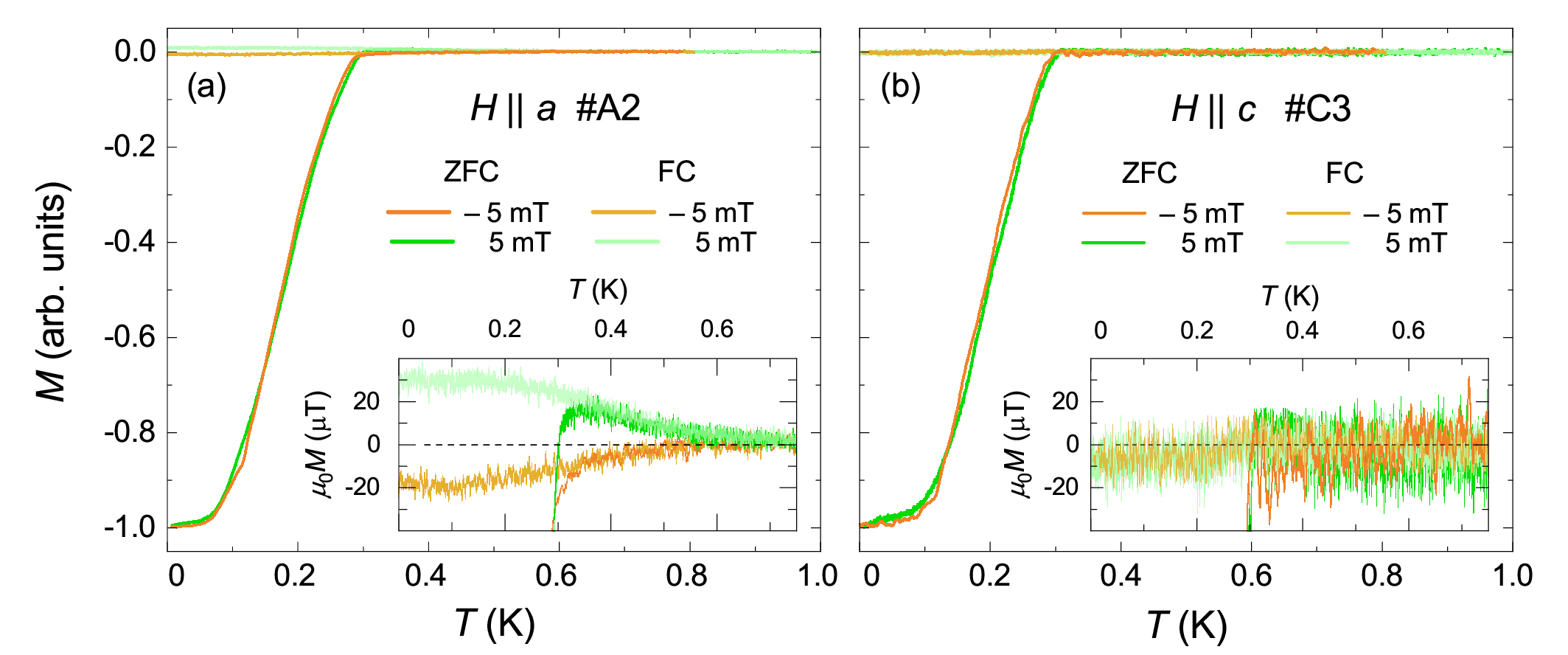}
		\caption{The in-plane (a) and out-of-plane (b) magnetization under the zero-field-cooling (ZFC) and field-cooling (FC) conditions measured in $\mu_0H$\,=\,5\,mT with the opposite directions.  Here, we assumed the ZFC data to represent the complete diamagnetic shielding at low temperatures. Insets highlight qualitative differences between the $M(T)$ results obtained in $H$\,$\parallel$\,$a$ and $H$\,$\parallel$\,$c$. For the in-plane magnetization, the amplitudes of the anomalous increase due to the $T_{0}$ phase are essentially the same for the ZFC and FC curves, but below the onset of superconductivity, the diamagnetic term dominates in the ZFC curves.}
		\label{}
	\end{figure*}

	\vspace{-5mm}
	\newpage
	\clearpage
	\subsection{VII. Evidence for the presence of zero-field staggered magnetization in a lower quality single crystal of \CRA.}

	\begin{figure*}[h]
		\centering
		\includegraphics[width=180mm,keepaspectratio=true]{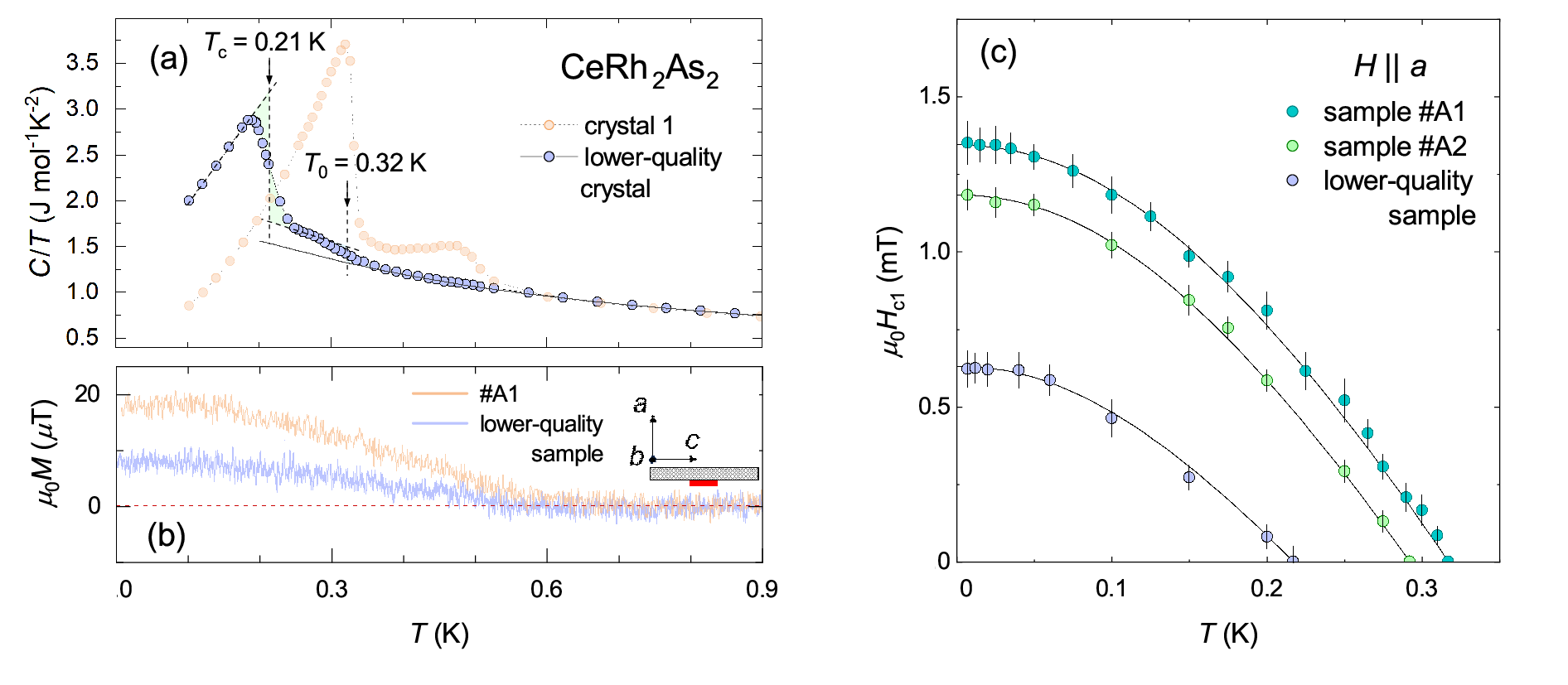}
		\caption{(a) Specific heat of a lower quality single crystal of \CRA, plotted as $C/T$ versus temperature, showing a rather broad superconducting jump at $T_c$ as low as 0.21\,K and lacking a clear signature of the $T_0$ phase. From the equal entropy construction we estimate $T_0$\,$\approx$\,0.32\,K. The transparent orange points show the $C/T$ results for a high-quality single crystal of \CRA\ with sharp signatures of superconductivity and undetermined $T_0$ phase. Interestingly, the differences between the two curves are negligible above $\approx$\,0.6\,K. (b) Similarities in the zero-field local magnetization of lower- and higher-quality samples. The $M(T)$ data were taken when the $a$ axis of each sample was perpendicular to the active area of Hall sensor. As in other $M(T)$ experiments discussed in the main text, the injection current of 1\,$\mu$A was used. Although the $T_0$ phase is barely visible in the heat capacity, the associated anomaly in the $M(T)$ data of lower-quality sample is quite similar to that observed in sample $\#$A1. It is also noteworthy that the temperature at which the increase in the in-plane magnetization occurs appears to be only weakly dependent on sample quality. (c) Temperature dependence of the lower critical field $H_{c 1}(T)$ along the $a$ axis of \CRA\ measured for samples with different $T_c$'s. Independently on sample quality, all sets of the $H_{c1}$($T$) data follow the conventional relation driven from the BCS theory (solid black lines).}
		\label{}
	\end{figure*}

	\clearpage
	\newpage
	
	\subsection{VIII. Superfluid density of \CRA\ and two-band model analysis.}

	\begin{figure*}[h]
		\centering
		\includegraphics[width=160mm,keepaspectratio=true]{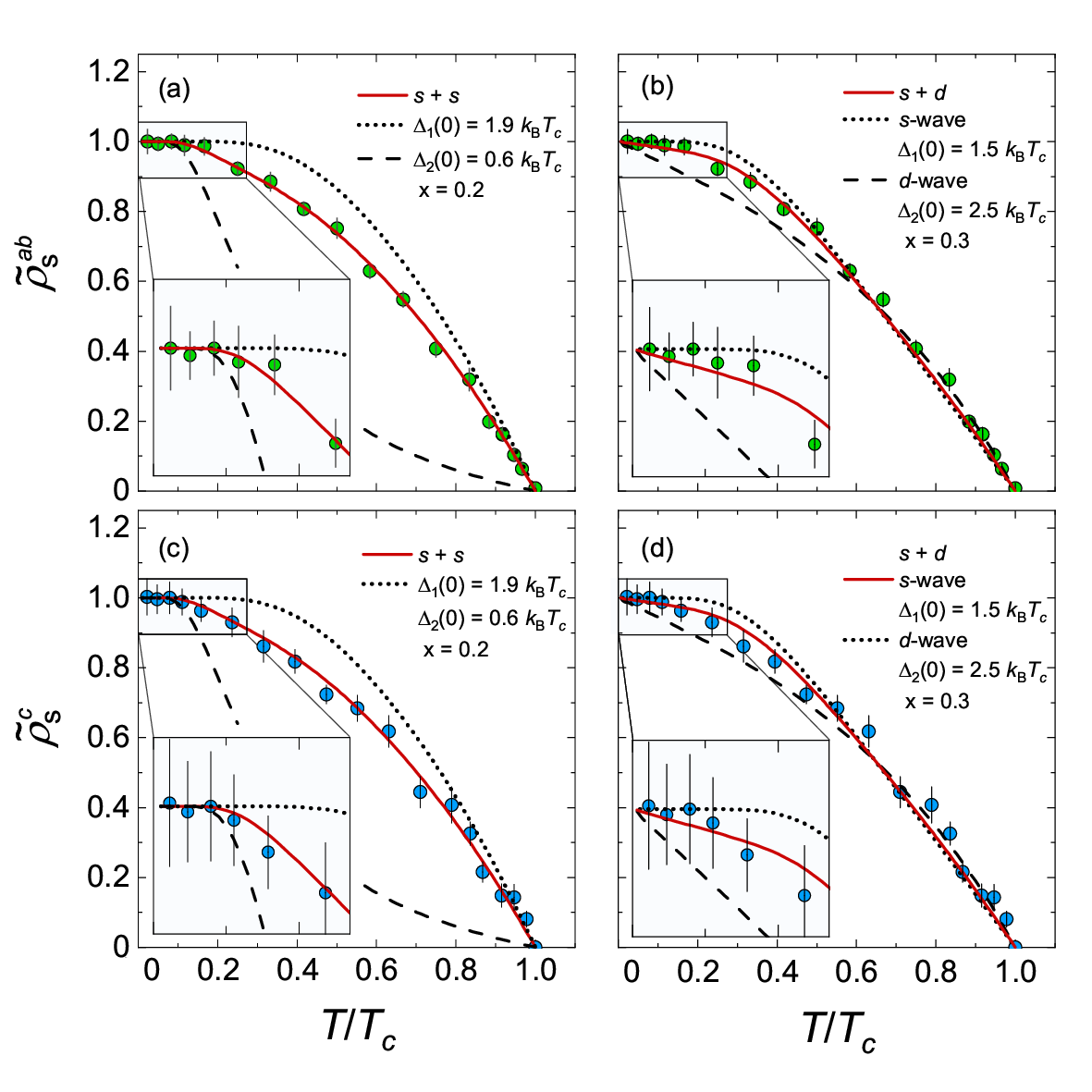}
		\caption{(a,b) Temperature dependence of the in-plane component of the superﬂuid density $\tilde{\rho}_{s}^{ab}$($T$) of CeRh$_{2}$As$_{2}$ derived from the $H_{c1}$$(T)$ data for $H$\,$\parallel$\,$c$ (green points) and fitted with different two-gap models (solid red lines). In each graph, the dotted black line shows $\tilde{\rho}_{s}$($T$) due to the larger gap $\Delta_{1}$ with the $s$-wave symmetry, while the dashed black line corresponds to the smaller gap $\Delta_{2}$ with a fraction $x$. While $\Delta_{2}$ has the $s$-wave symmetry in (a), (b) shows the case of the $s\,+\,d$ model. (c,d) Similar analysis but for the out-of-plane superﬂuid density $\tilde{\rho}_{s}^{c}$($T$) derived from the combined $H_{c1}$$(T)$ data for $H$\,$\parallel$\,$a$ and $H$\,$\parallel$\,$c$ (cf., SM v). Insets: zoom into the temperature range below 0.25$T_{c}$.Insets: Zoom in on the temperature range below 0.25$T_{c}$. Note the difference between the $s$\,+\,$s$ and $s$\,+\,$d$ fits. The $s$\,+\,$s$ model describes well the flat temperature dependence below 0.1$T_{c}$, while the $s$\,+\,$d$ model is linear down to the lowest temperatures.}
		\label{}
	\end{figure*}


\begin{thebibliography}{99}
	
	\bibitem{Sato} M. Sato and Y. Ando, Topological superconductors: A\,review, Rep. Prog. Phys. 80, 076501 (2017).
	
	\bibitem{Sigrist} M. Sigrist and K. Ueda, Phenomenological theory of unconventional superconductivity, Rev. Mod. Phys. 63, 239 (1991).
		
	\bibitem{Khim2021} S. Khim, J. F. Landaeta, J. Banda, N. Bannor, M.\,Brando, P. M. R. Brydon, D. Hafner, R. Küchler, R.\,Cardoso-Gil, U.	Stockert, A. P. Mackenzie, D.\,F.\,Agterberg, C. Geibel, and E. Hassinger, Field-induced transition within the superconducting state of CeRh$_{2}$As$_{2}$ Science 373, 1012 (2021).
	
	\bibitem{Landaeta2022}  J. F. Landaeta, P. Khanenko, D. C. Cavanagh, C.\,Geibel, S.	Khim, S. Mishra, I. Sheikin, P. M. R. Brydon, D.\,F.\,Agterberg, M. Brando, and E. Hassinger, Field-angle dependence reveals odd-parity superconductivity in CeRh$_{2}$As$_{2}$, Phys. Rev. X 12, 031001 (2022).
	
	\bibitem{Schertenleib2021} E. G. Schertenleib, M. H. Fischer, and M. Sigrist, Unusual $H$ - $T$ phase diagram of CeRh$_{2}$As$_{2}$: The role of staggered noncentrosymmetricity, Phys. Rev. Research 3, 023179  (2021).
	
	\bibitem{Mockli2021} D. Möckli and A. Ramires, Two scenarios for superconductivity in CeRh$_{2}$As$_{2}$, Phys. Rev. Research 3, 023204  (2021).
	
	\bibitem{Nogaki2021} K. Nogaki, A. Daido, J. Ishizuka, and Y. Yanase, Topological crystalline superconductivity in locally noncentrosymmetric CeRh$_{2}$As$_{2}$, Phys. Rev. Research 3, L032071 (2021).
	
	\bibitem{Cavanagh2022} D. C. Cavanagh, T. Shishidou, M. Weinert, P. M. R. Brydon, and Daniel F. Agterberg, Nonsymmorphic symmetry and field-driven odd-parity pairing in CeRh$_{2}$As$_{2}$, Phys. Rev. B 105, L020505 (2022).
	
	\bibitem{Nogaki2022} K. Nogaki, and Y. Yanase, Even-odd parity transition in strongly correlated locally noncentrosymmetric superconductors: Application to CeRh$_{2}$As$_{2}$, Phys. Rev. B 106, L100504 (2022).
	
	\bibitem{Suh2023} H. G. Suh, Y. Yu, T. Shishidou, M. Weinert, P. Brydon, and D. F. Agterberg, Superconductivity of anomalous pseudospin in nonsymmorphic materials, Phys. Rev. Res. 5, 033204 (2023).
	
	\bibitem{Szabo2023} A. L. Szabó, M. H. Fischer, and M. Sigrist, Effects of
	nucleation at a first-order transition between two superconducting phases: Application to CeRh$_{2}$As$_{2}$, Phys. Rev. Res. 6, 023080 (2024).
	
	\bibitem{Lee2024} C. Lee, D. F. Agterberg, and P. M. R. Brydon, Unified picture of superconductivity and magnetism in CeRh$_{2}$As$_{2}$, arXiv:2407.00536 (2024).
	
	\bibitem{Ishizuka2024} J. Ishizuka, K. Nogaki, M.  Sigrist, and Y. Yanase, Correlation-induced Fermi surface evolution and topological crystalline superconductivity in CeRh$_{2}$As$_{2}$, Phys. Rev. B 110, L140505 (2024).
	
	\bibitem{Machida2022} K. Machida, Violation of Pauli-Clogston limit in the heavy-fermion superconductor  CeRh$_{2}$As$_{2}$: Duality of itinerant and localized 4$f$ electrons, Phys. Rev. B 106, 184509 (2022).
	
	\bibitem{Hazra2023}  T. Hazra and P. Coleman, Triplet pairing mechanisms
	from Hund’s-Kondo models: Applications to UTe$_{2}$ and CeRh$_{2}$As$_{2}$, Phys. Rev. Lett. 130, 136002 (2023).
	
	\bibitem{Ogata2023} S. Ogata, S. Kitagawa, K. Kinjo, K. Ishida, M. Brando, E. Hassinger, C. Geibel, and S. Khim, Parity transition of spin-singlet superconductivity using sublattice degrees of freedom, Phys. Rev. Lett. 130, 166001 (2023).
	
	\bibitem{Kitagawa2022} S. Kitagawa, M. Kibune, K. Kinjo, M. Manago, T.\,Taniguchi, K. Ishida, M. Brando, E. Hassinger, C.\,Geibel, and S. Khim,
	Two-dimensional XY-type magnetic properties of locally noncentrosymmetric superconductor CeRh$_{2}$As$_{2}$, J. Phys. Soc. Jpn. 91, 043702 (2022).
	
	\bibitem{Ogata2024} S. Ogata, S. Kitagawa, K. Kinjo, K, Ishida, M. Brando, E. Hassinger, C. Geibel, and S. Khim, Appearance of $c$-axis magnetic moment in odd-parity antiferromagnetic
	state in CeRh$_{2}$As$_{2}$ revealed by $^{75}$As-NMR, Phys. Rev. B 110, 214509 (2024).
	
	
	\bibitem{Kibune2022} M. Kibune, S. Kitagawa, K. Kinjo, S. Ogata, M. Manago, T. 	Taniguchi, K. Ishida, M. Brando, E. Hassinger, H. Rosner, C. Geibel, and S. Khim, Observation of antiferromagnetic order as odd-parity multipoles inside the superconducting phase in CeRh$_{2}$As$_{2}$, Phys. Rev. Lett. 128, 057002 (2022).
	
	\bibitem{Chajewski2} G. Chajewski and D. Kaczorowski, Discovery of magnetic phase transitions in heavy-fermion superconductor CeRh$_{2}$As$_{2}$, Phys. Rev. Lett. 132, 076504 (2024).
	
	\bibitem{Onishi2022} S. Onishi, U. Stockert, S. Khim, J. Banda, M. Brando, E. Hassinger, 
	Low-temperature thermal conductivity of the two-phase superconductor CeRh$_{2}$As$_{2}$,  Front. Electron. Mater. 2, 880579 (2022).
	
	\bibitem{Siddiquee2023} H. Siddiquee, Z. Rehfuss, C. Broyles, and S. Ran, Pressure dependence of superconductivity in CeRh$_{2}$As$_{2}$, Phys. Rev. B 108, L020504 (2023).
	
	
	\bibitem{Hafner2022}  D. Hafner, P. Khanenko, E.-O. Eljaouhari, R. Küchler, J. 	Banda, N. Bannor, T. Lühmann, J. F. Landaeta, S. Mishra, I. 	Sheikin, E. Hassinger, S. Khim, C. Geibel, G.\,Zwicknagl, and M. Brando, Possible quadrupole density wave in the superconducting Kondo lattice CeRh$_{2}$As$_{2}$, Phys. Rev. X 12, 011023 (2022).
	

	\bibitem{Christovam2024} D. S. Christovam, M. Ferreira-Carvalho, A. Marino, M. Sundermann, D. Takegami, A. Melendez-Sans, K. D. Tsuei, Z. Hu, S. Rößler, M. Valvidares, M. W. Haverkort, Y. Liu, E. D. Bauer, L. H. Tjeng, G. Zwicknagl, and A. Severing, Spectroscopic evidence of Kondo-induced quasiquartet in CeRh$_{2}$As$_{2}$, Phys. Rev. Lett. 132, 046401 (2024).
	
	\bibitem{Khim2024} S. Khim, O. Stockert, M. Brando, C. Geibel, C. Baines, T. J. Hicken, H. Luetkens, D. Das, T. Shiroka, Z.\,Guguchia, and R. Scheuermann, Coexistence of local magnetism and superconductivity in the heavy-fermion CeRh$_{2}$As$_{2}$ revealed by $\mu$SR studies, arXiv:2406.16575 (2024).
	
	\bibitem{Chen2024arxiv} T. Chen, H. Siddiquee, Z. Rehfuss, S. Gao, C. Lygouras, J. Drouin, V. Morano, K. E. Avers, C.\,J.\,Schmitt, A. Podlesnyak, S. Ran, Y. Song, and C. Broholm, Quasi-two-dimensional antiferromagnetic spin fluctuations in the spin-triplet superconductor candidate CeRh$_{2}$As$_{2}$, arXiv:2406.03566 (2024).
	
	\bibitem{Schmidt2024} B. Schmidt and P. Thalmeier, Anisotropic magnetic and quadrupolar $H$-$T$ phase diagram of CeRh$_{2}$As$_{2}$, Phys. Rev. B 110, 075154 (2024).
	
	\bibitem{Thalmeier2024} P. Thalmeier, A. Akbari, and B. Schmidt, Thermodynamics, elastic anomalies and excitations in the field induced phases of CeRh$_{2}$As$_{2}$, arxiv:2412.02537 (2024).
	
	\bibitem{Chajewski1} G. Chajewski, D. Szymański, M. Daszkiewicz, and D. Kaczorowski, Horizontal flux growth as an efficient preparation method of CeRh$_{2}$As$_{2}$ single crystals, Mater. Horiz. 11, 855 (2024).
	
	\bibitem{SM} See Supplemental Material at [URL] which includes Refs.
	\cite{Chajewski1,Khim2021,Kim2019,Min2017,Li2019,Okazaki2010,Abdel2013,Cieplak2013,Yamashita2017,Collignon2017,Juraszek2020,Ishihara2023,Juraszek2024,Joshi2019,Tinkham,Szabo2023,Ogata2023,Takenaka2017}, for details on single crystal growth and characterization,  advantages of micro-Hall probe magnetometry, estimation of the field of first flux penetration and corresponding $H_{c1}$ value, intrinsic origin of the saturation of $H_{c1}$($T$) in the $T$\,=\,0 limit, parameters characterizing the SC1 derived from lower critical field measurements, sign reversal of the in-plane magnetization at low magnetic fields, evidence for the presence of zero-field staggered magnetization in a lower quality single crystal of
	CeRh$_{2}$As$_{2}$, and superfluid density of \CRA\ and two-band model analysis.
	
	
\bibitem{Kim2019}	M. Kim, P. Kumaravadivel, J. Birkbeck, W. Kuang, S. G. Xu, D. G. Hopkinson, J. Knolle, P. A. McClarty, A. I. Berdyugin, M. Ben Shalom, R. V. Gorbachev, S. J. Haigh, S. Liu, J. H. Edgar, K. S. Novoselov, I. V. Grigorieva, and A. K. Geim, Micromagnetometry of two-dimensional ferromagnets, Nat. Electron 2, 457 (2019).

	\bibitem{Min2017} C. Min , J. Park, J. K. Mun, Y. Lim, J. Min, J.-W. Lim, D.M. Kang, H.K. Ahn, T.H. Shin, J. Cheon, H.S. Lee, R. Weissleder, C. M. Castro, H. Lee, Integrated microHall magnetometer to measure the magnetic properties of nanoparticles, Lab Chip, 17, 4000 (2017).
	
	\bibitem{Li2019} X. Li, C. Collignon, L. Xu, H. Zuo, A. Cavanna, U. Gennser, D. Mailly, B. Fauqué, L. Balents, Z. Zhu, and K. Behnia, Chiral domain walls of Mn$_{3}$Sn and their memory. Nat Commun 10, 3021 (2019).
	
	\bibitem{Okazaki2010} R. Okazaki, M. Shimozawa, H. Shishido, M. Konczykowski, Y. Haga, T. D. Matsuda, E. Yamamoto, Y. Onuki, Y. Yanase, T. Shibauchi, Y. Matsuda, Anomalous Temperature Dependence of Lower Critical Field in Ultraclean URu$_{2}$Si$_{2}$, J. Phys. Soc. Jpn. 79, 084705 (2010)
	
	\bibitem{Abdel2013} M. Abdel-Hafiez, J. Ge, A. N. Vasiliev, D. A. Chareev, J. Van de Vondel, V. V. Moshchalkov, and A. V. Silhanek, Temperature dependence of lower critical field $H_{c1}$($T$) shows nodeless superconductivity in FeSe, Phys. Rev. B 88, 174512 (2013).
	
	\bibitem{Cieplak2013} M. Z. Cieplak, Z. Adamus, M. Kończykowski, L. Y. Zhu,, X. M. Cheng, and C. L. Chien, Tuning vortex confinement by magnetic domains in a superconductor/ferromagnet bilayer, Phys. Rev. B 87, 014519 (2013).
	
	\bibitem{Yamashita2017}  T. Yamashita, T. Takenaka, Y. Tokiwa, J. A. Wilcox, Y. Mizukami, D. Terazawa, Y. Kasahara, S. Kittaka, T. Sakakibara, M. Konczykowski, S. Seiro, H. S. Jeevan, C. Geibel, C. Putzke, T. Onishi, H. Ikeda, A. Carrington, T. Shibauchi, and Y. Matsuda, Fully gapped superconductivity with no sign change in the prototypical heavy-fermion CeCu$_{2}$Si$_{2}$,
	Sci. Adv. 3, e1601667 (2017).
	
	\bibitem{Collignon2017} C. Collignon, B. Fauqué, A. Cavanna, U. Gennser, D. Mailly, and K. Behnia, Superfluid density and carrier concentration across a superconducting dome: The case of strontium titanate, Phys. Rev. B 96, 224506 (2017).
	
		
	\bibitem{Ishihara2023} K. Ishihara, M. Kobayashi, K. Imamura, M. Konczykowski, H. Sakai, P. Opletal, Y. tokiwa, Y. Haga, K. Hashimoto, and T. Shibauchi, Anisotropic enhancement of lower critical field inultraclean crystals of spin- triplet superconductor candidate UTe$_{2}$. Phys. Rev. Res. 5, l022002 (2023).
	
	\bibitem{Juraszek2024} J. Juraszek, Yu. V. Sharlai, M. Konczykowski, A. Ślebarski, and T. Cichorek, Two-band superconductivity with weak interband coupling in structurally disordered Y$_{5}$Rh$_{6}$Sn$_{18}$. Phys. Rev. B 109, 174526 (2024).
	
	\bibitem{Joshi2019} K. R. Joshi, N. M. Nusran, M. A. Tanatar, K. Cho, W. R. Meier, S. L. Bud’ko, P. C. Canfield, and R. Prozorov, Measuring the Lower Critical Field of Superconductors Using Nitrogen-Vacancy Centers in Diamond Optical Magnetometry, Phys. Rev. Applied 11, 014035 (2019).
	
	\bibitem{Tinkham} M. Tinkham, Introduction to Superconductivity (Krieger, Malabar, Florida, 1975).
	
	
	\bibitem{Takenaka2017}	T. Takenaka, Y. Mizukami, J. A. Wilcox, M. Konczykowski, S. Seiro, C. Geibel, Y. Tokiwa, Y. Kasahara, C. Putzke, Y. Matsuda, A. Carrington, and T. Shibauchi, Full-gap superconductivity robust against disorder in heavy-fermion, Phys. Rev. Lett. 119, 077001 (2017).
	
	
	\bibitem{Juraszek2020} J. Juraszek, R. Wawryk, Z. Henkie, M. Konczykowski, and T. Cichorek, Symmetry of order parameters in multiband superconductors LaRu$_{4}$As$_{12}$ and PrOs$_{4}$As$_{12}$ probed by local magnetization measurements, Phys. Rev. Lett. 124, 027001 (2020).
	
	\bibitem{Ren2008} C. Ren, Z.-S. Wang, H.-Q. Luo, H. Yang, L. Shan, and	H.-H. Wen, Evidence for two energy gaps in superconducting Ba$_{0.6}$K$_{0.4}$Fe$_{2}$As$_{2}$ single crystals and the breakdown of the Uemura plot, Phys. Rev. Lett. 101, 257006 (2020).
	
	\bibitem{Ge2012} J. Ge, J. Gutierrez, B. Raes, T. Watanabe, J. Koshio, and V.V. Moshchalkov, Two energy gaps in superconducting Lu$_{2}$Fe$_{3}$Si$_{5}$ single crystal derived from the temperature dependence of lower critical field $H_{c1}$($T$), Physica C 478, 5 (2012).
	
	\bibitem{Hafiez2013} M. Abdel-Hafiez, J. Ge, A. N. Vasiliev, D. A. Chareev, J.\,Van de Vondel, V. V. Moshchalkov, and A. V. Silhanek, Temperature dependence of lower critical field $H_{c1}$($T$) shows nodeless superconductivity in FeSe, Phys. Rev. B 88, 174512 (2013).
	
	
	\bibitem{Kittaka2014} S. Kittaka, Y. Aoki, Y. Shimura, T. Sakakibara, S. Seiro, C.
	Geibel, F. Steglich, H. Ikeda, and K. Machida, Multiband superconductivity with unexpected deficiency of nodal quasiparticles in CeCu$_{2}$Si$_{2}$, Phys. Rev. Lett. 112, 067002 (2014).

	\bibitem{Pang2017} G. Pang, M. Smidman, J. Zhang, L. Jiao, Z. Weng, E.M. Nica, Y. Chen, W. Jiang, Y. Zhang, W. Xie, H.S. Jeevan, H. Lee, P. Gegenwart, F. Steglich, Q. Si, and H. Yuan, Fully gapped d-wave superconductivity in CeCu$_{2}$Si$_{2}$, Proc. Natl. Acad. Sci. U.S.A.
	115, 5343 (2017).
		
	\bibitem{Chen2024} X. Chen, L. Wang, J. Ishizuka, R. Zhang, K. Nogaki, Y. Cheng, F. Yang, Z. Chen, F. Zhu, Z. Liu, J. Mei, Y. Yanase, B. Lv, and Y. Huang, Coexistence of near-$E_{F}$ flat band and van Hove singularity in a two-phase superconductor, Phys. Rev. X 14, 021048 (2024).
	
	
	\bibitem{Cheong} S.-W. Cheong, M. Fiebig, W. Wu, L. Chapon, and V.\,Kiryukhin, Seeing is believing: visualization of antiferromagnetic domains, npj Quantum Mater. 5, 3 (2020).

	
	
	
	
\end{thebibliography}

\newpage

	
		
		
		\begin{thebibliography}{}
			\bibitem{Chajewski1}  G. Chajewski, D. Szymański, M. Daszkiewicz, and D. Kaczorowski, Horizontal flux growth as an efficient preparation method of CeRh$_{2}$As$_{2}$ single crystals, Mater. Horiz. 11, 855 (2024).
			
			\bibitem{Khim2021} S. Khim, J. F. Landaeta, J. Banda, N. Bannor, M. Brando, P. M. R. Brydon, D. Hafner, R. Küchler, R.\,Cardoso-Gil, U.	Stockert, A. P. Mackenzie, D. F. Agterberg, C. Geibel, and E. Hassinger, Field-induced transition within the superconducting state of CeRh$_{2}$As$_{2}$ Science 373, 1012 (2021).
			
			
			\bibitem{Kim2019}	M. Kim, P. Kumaravadivel, J. Birkbeck, W. Kuang, S. G. Xu, D. G. Hopkinson, J. Knolle, P. A. McClarty, A. I. Berdyugin, M. Ben Shalom, R. V. Gorbachev, S. J. Haigh, S. Liu, J. H. Edgar, K. S. Novoselov, I. V. Grigorieva, and A. K. Geim, Micromagnetometry of two-dimensional ferromagnets, Nat. Electron 2, 457 (2019).
			
			\bibitem{Min2017} C. Min , J. Park, J. K. Mun, Y. Lim, J. Min, J.-W. Lim, D.M. Kang, H.K. Ahn, T.H. Shin, J. Cheon, H.S. Lee, R. Weissleder, C. M. Castro, H. Lee, Integrated microHall magnetometer to measure the magnetic properties of nanoparticles, Lab Chip, 17, 4000 (2017).
			
			\bibitem{Li2019} X. Li, C. Collignon, L. Xu, H. Zuo, A. Cavanna, U. Gennser, D. Mailly, B. Fauqué, L. Balents, Z. Zhu, and K. Behnia, Chiral domain walls of Mn$_{3}$Sn and their memory. Nat Commun 10, 3021 (2019).
			
			\bibitem{Okazaki2010} R. Okazaki, M. Shimozawa, H. Shishido, M. Konczykowski, Y. Haga, T. D. Matsuda, E. Yamamoto, Y. Onuki, Y. Yanase, T. Shibauchi, Y. Matsuda, Anomalous Temperature Dependence of Lower Critical Field in Ultraclean URu$_{2}$Si$_{2}$, J. Phys. Soc. Jpn. 79, 084705 (2010)
			
			\bibitem{Abdel2013} M. Abdel-Hafiez, J. Ge, A. N. Vasiliev, D. A. Chareev, J. Van de Vondel, V. V. Moshchalkov, and A. V. Silhanek, Temperature dependence of lower critical field $H_{c1}$($T$) shows nodeless superconductivity in FeSe, Phys. Rev. B 88, 174512 (2013).
			
			\bibitem{Cieplak2013} M. Z. Cieplak, Z. Adamus, M. Kończykowski, L. Y. Zhu,, X. M. Cheng, and C. L. Chien, Tuning vortex confinement by magnetic domains in a superconductor/ferromagnet bilayer, Phys. Rev. B 87, 014519 (2013).
			
			\bibitem{Yamashita2017}  T. Yamashita, T. Takenaka, Y. Tokiwa, J. A. Wilcox, Y. Mizukami, D. Terazawa, Y. Kasahara, S. Kittaka, T. Sakakibara, M. Konczykowski, S. Seiro, H. S. Jeevan, C. Geibel, C. Putzke, T. Onishi, H. Ikeda, A. Carrington, T. Shibauchi, and Y. Matsuda, Fully gapped superconductivity with no sign change in the prototypical heavy-fermion CeCu$_{2}$Si$_{2}$,
			Sci. Adv. 3, e1601667 (2017).
			
			\bibitem{Collignon2017} C. Collignon, B. Fauqué, A. Cavanna, U. Gennser, D. Mailly, and K. Behnia, Superfluid density and carrier concentration across a superconducting dome: The case of strontium titanate, Phys. Rev. B 96, 224506 (2017).
			
			
			\bibitem{Takenaka2017} T. Takenaka, Y. Mizukami, J. A. Wilcox, M. Konczykowski, S. Seiro, C. Geibel, Y. Tokiwa, Y. Kasahara, C. Putzke, Y.
			Matsuda, A. Carrington, and T. Shibauchi, Full-Gap Superconductivity Robust against Disorder in Heavy-Fermion CeCu$_{2}$Si$_{2}$, Phys. Rev. Lett. 119, 077001 (2017).
			
			\bibitem{Juraszek2020} J. Juraszek, R. Wawryk, Z. Henkie, M. Konczykowski, and T. Cichorek, Symmetry of Order Parameters in Multiband Superconductors LaRu$_{4}$As$_{12}$ and PrOs$_{4}$As$_{12}$ Probed by Local Magnetization Measurements, Phys. Rev. Lett. 124, 027001 (2020).
			
			\bibitem{Ishihara2023} K. Ishihara, M. Kobayashi, K. Imamura, M. Konczykowski, H. Sakai, P. Opletal, Y. tokiwa, Y. Haga, K. Hashimoto, and T. Shibauchi, Anisotropic enhancement of lower critical field inultraclean crystals of spin- triplet superconductor candidate UTe$_{2}$. Phys. Rev. Res. 5, l022002 (2023).
			
			\bibitem{Juraszek2024} J. Juraszek, Yu. V. Sharlai, M. Konczykowski, A. Ślebarski, and T. Cichorek, Two-band superconductivity with weak interband coupling in structurally disordered Y$_{5}$Rh$_{6}$Sn$_{18}$. Phys. Rev. B 109, 174526 (2024).
			
			\bibitem{Joshi2019} K. R. Joshi, N. M. Nusran, M. A. Tanatar, K. Cho, W. R. Meier, S. L. Bud’ko, P. C. Canfield, and R. Prozorov, Measuring the Lower Critical Field of Superconductors Using Nitrogen-Vacancy Centers in Diamond Optical Magnetometry, Phys. Rev. Applied 11, 014035 (2019).
			
			\bibitem{Tinkham} M. Tinkham, Introduction to Superconductivity (Krieger, Malabar, Florida, 1975).
			
			\bibitem{Szabo2023} A. L. Szabó, M. H. Fischer, and M. Sigrist, Effects of
			nucleation at a first-order transition between two superconducting phases: Application to CeRh$_{2}$As$_{2}$, Phys. Rev. Res. 6, 023080 (2024).
			
			\bibitem{Ogata2023} S. Ogata, S. Kitagawa, K. Kinjo, K. Ishida, M. Brando, E. Hassinger, C. Geibel, and S. Khim, Parity transition of spin-singlet superconductivity using sublattice degrees of freedom, Phys. Rev. Lett. 130, 166001 (2023).
			
		
			
			
		
			
			
		\end{thebibliography}
		\end{document}